\documentclass[12pt]{article}
\usepackage{amsmath}
\usepackage{graphicx,psfrag,epsf}
\usepackage{enumerate}
\usepackage{url} 

\usepackage{array}

\usepackage{amsthm}

\usepackage{diagbox}

\usepackage[T1]{fontenc}

\usepackage[toc,page]{appendix}

\usepackage{epstopdf}
\usepackage[caption=false]{subfig}

\usepackage[numbers,sort&compress]{natbib}
\bibpunct[, ]{[}{]}{,}{n}{,}{,}

\newtheorem{theorem}{Theorem}[section]

\theoremstyle{definition}

\newtheorem{definition}[theorem]{Definition}

\usepackage[english]{babel}
\usepackage{amssymb}
\usepackage{textcmds}
\usepackage{hyperref}

\usepackage{xcolor}

\usepackage{graphicx}
\usepackage[section]{placeins}

\usepackage{float}

\usepackage{url}

\usepackage{breakurl}

\usepackage{float}

\usepackage[caption = false]{subfig}
\usepackage{authblk}

\newcommand{\blind}{0}

\begin{document}

\def\spacingset#1{\renewcommand{\baselinestretch}%
{#1}\small\normalsize} \spacingset{1}


\if0\blind
{
  \title{\bf Statistical comparison of quality attributes: a range-based approach}

\author[1]{Gerhard G\"ossler\thanks{
    Corresponding author: gerhard.goessler@uni-graz.at}}
    \author[2]{Vera Hofer}
\author[3]{Hans Manner}
\author[1]{Walter Goessler}

\affil[1]{Department of Analytical Chemistry, University of Graz}
\affil[2]{Department of Operations Research, University of Graz}
\affil[3]{Department of Economics, University of Graz}

  \maketitle
} \fi

\if1\blind
{
  \bigskip
  \bigskip
  \bigskip
  \begin{center}
    {\LARGE\bf Title}
\end{center}
  \medskip
} \fi

\vspace*{-1\baselineskip}

\bigskip
\begin{abstract}
\noindent A novel approach for comparing quality attributes of different products when there is considerable product-related variability is proposed. In such a case, the whole range of possible realizations must be considered. Looking, for example, at the respective information published by agencies like the \label{EF} EMA or the FDA, one can see that commonly accepted tests together with the proper statistical framework are not yet available. This work attempts to close this gap in the  treatment of range-based comparisons. The question of when two products can be considered similar with respect to a certain property is discussed and a  framework for such a statistical comparison is presented, which is based on the proposed concept of $\kappa$-cover. Assuming normally distributed quality attributes a statistical test termed covering-test is proposed. Simulations show that this test  possesses desirable statistical properties with respect to small sample size and power. In order to demonstrate the usefulness of the suggested concept, the proposed test is applied to a data set from the pharmaceutical industry. 
\end{abstract}

\noindent%
\vfill

\newpage
\spacingset{1} 


\maketitle

\section{Introduction}

Comparing quality characteristics of drug products is of fundamental significance in drug development and production. For example, when developing a generic drug, one is striving for therapeutic equivalence to a reference medicinal product (originator) already on the market. For receiving approval by the market authority, one of the hurdles that have to be overcome is to demonstrate the similarity of the two products with respect to several quality attributes \label{QA} (QAs). Such QAs of interest can be, e.g., the impurity profile, the zeta potential or the polydispersity index. If possible, statistical methods for evaluation of similarity should be used as discussed in the EMA reflection paper on this topic\cite{EMA.2021}. Unfortunately, there is no \qq{one-fits-all approach} since the characteristics of the underlying data generating processes can differ between QAs and products with respect to scale of measurement, sources of variability and distribution properties. Also additional information regarding specified values and tolerances has to be integrated into the statistical analysis if available.

The choice of the statistical approach is largely determined by the sources of variability in the analytical data. If the variability in the data can solely be attributed to measurement error, comparing the mean values only is sufficient. This can be done by applying a two-sample t-test. If information about the specified value and tolerated deviations is available, it is sufficient to compare the mean value with the range given thereby by applying a one-sample t-test for equivalence. Two-sample equivalence tests like the two one sided t-tests procedure \label{TOST} (TOST~\cite{Schuirmann.1987}) are of interest only in the rather special case where the tolerated deviations are known but the specified value is not.\\ 

However, in the case of product-related variability one is naturally interested in the ranges\cite{EMA.2021}\cite{Mielke.2018} of the values of critical QAs in order to ensure that a compliant product is manufactured. In our understanding, this is best done by comparing the respective distributions or their quantiles. In the following, these distributions will be denoted reference and test distribution, $D_R$ and $D_T$, and the statistical approach for the comparison of $D_T$ with $D_R$ 
presented in this work is inspired by applications where knowledge about specified values and tolerated deviations is missing as is often the case in the development of generic drugs. All available information is obtained from the (chemical, physical, $\ldots$) characterization of suitable samples taken from several batches of the products to be compared. Due to this, the assessment of QAs is often referred to as the assessment of \textit{analytical similarity}\cite{ChungChow.2014}. This term is also used in the remainder of this paper. Since it seems reasonable to assume that the product already on the market is within the unknown specifications, the data obtained from the analysis of some reference batches should make it possible to determine reasonable bounds, which are well within the unknown accepted ranges. Therefore, the methodology proposed in this work is based on the following assumption:\\

A test product $T$ can be considered \textbf{analytically similar} \textit{in the presence of product related variability} to its corresponding reference product $R$ with respect to a selected QA, if the range of $D_{R}$ \emph{suitably covers} the range of $D_{T}$.\\

Defining what is meant by \qq{$D_{R}$ suitably covers the range of $D_{T}$} is of course unavoidable when attempting a mathematical treatment of the problem. 
Since it seems reasonable to allow some scope for $D_{T}$ within the range of $D_{R}$, total equality of the distributions at hand is just a special case in this context.  The range of a distribution can be characterized by an appropriate combination of information regarding scale and location, so comparing the two distributions with respect to expected values and/or variances separately will yield an insufficient answer to this problem. No generally accepted theoretical framework for this kind of comparison exists (see, e.g., \cite{EMA.2021}) the approach proposed in this work is an attempt to close this gap. To this end, a new way of comparing two distributions with respect to their ranges is proposed. It is based on the concept of $\kappa$-cover we introduce below. Such a concept is also a necessary prerequisite for the construction of proper statistical tests. We introduce such a test in this paper.\\

Since in the pharmaceutical industry comparing QAs of drugs is essential, it is hardly surprising that several different approaches for testing of analytical similarity in the presence of product related variability have already been developed by researchers and governmental authorities. However, there is debate as to whether these concepts are really suitable to manage the needed range-based comparisons properly. Furthermore, these approaches are often designed to enable a relatively large deviation of $D_T$ from $D_R$, which is something the approach proposed in this work is explicitly not intended for. Approaches designed for allowing pre-specified tolerance might therefore have to be adapted accordingly in order to be applicable in cases where no knowledge about possible deviations is available.\\ 

In the context of biosimilars, the FDA released two guidances which contain approaches for the comparison of QAs when product related variability is present. The first, suggested in 2017\cite{FDA.2017} was withdrawn in 2018\cite{FDA.2018} after the agency received considerable criticism from the public. The tests were, depending on the relevance of the QA to clinical outcomes, a reference scaled equivalence test\cite{Weng.2019}, a quality range (\label{QR} \textit{QR}) approach, and an assessment based on a graphical comparison. In 2019, the FDA released a successive draft guidance\cite{FDA.2019} in which the equivalence test is not mentioned anymore. Several publications discuss the FDA approaches; see, e.g., the work of Tsong et al.\cite{Tsong.2017}, Chow et al.\cite{Chow.2016}, Wang \& Chow\cite{Wang.2017} and  Son et al.\cite{Son.2020}. The criticism of such approaches is considerable as summarized in the work of Chow \& Lee\cite{Chow.2021}. Generally speaking, the objections raised against these approaches criticize a lack of scientific/statistical justification. As the FDA has to date not resolved such critical issues (see again Chow \& Lee\cite{Chow.2021}), a scientific basis for agreement remains lacking. However, the approaches suggested by the FDA, and especially the problems associated with them, are a clear sign of the complexity of the comparability issue.
Other proposed QR-approaches for comparing ranges such as the min/max and the tolerance interval approach differ mainly with respect to the calculation of the QR and have similar disadvantages\cite{EMA.2021} as the QR approach proposed by the FDA. In order to overcome the problems entailed in using the aforementioned QR approaches, Boulanger\cite{Boulanger.2016} suggested testing whether the range of $D_{R}$ covers the range of $D_{T}$ by using a certain tolerance interval ($\beta/\gamma-$TI) based on the observations from $D_{R}$, and a certain prediction interval ($\beta$-PI) based on that from $D_{T}$, i.e., one has to determine whether the TI fully includes the PI. One obvious problem with the approach proposed by Boulanger is that it is not possible to establish a proper relationship between any range-based hypotheses and the parametrization of the test, i.e., to give a formal definition of what is meant by the covering of one range by another, and hence no explicit hypotheses can be given. The tail test, suggested by Mielke et al.\cite{Mielke.2018}, deals also with biological drugs and constitutes an approach that also tries to compare distributions with respect to ranges, but differs significantly from the QR type approaches mentioned so far. Despite some resemblance to our concept of $\kappa$-cover with respect to the basic concept of a range based comparison of distributions, the approach of Mielke et al. differs largely compared to ours in many aspects rendering their approach not equivalent with respect to the underlying philosophy and also the resulting statistical framework, e.g., the tested hypotheses.\\

\begin{figure*}[h!]
 \centering

\includegraphics[width=12.5cm,  keepaspectratio,]{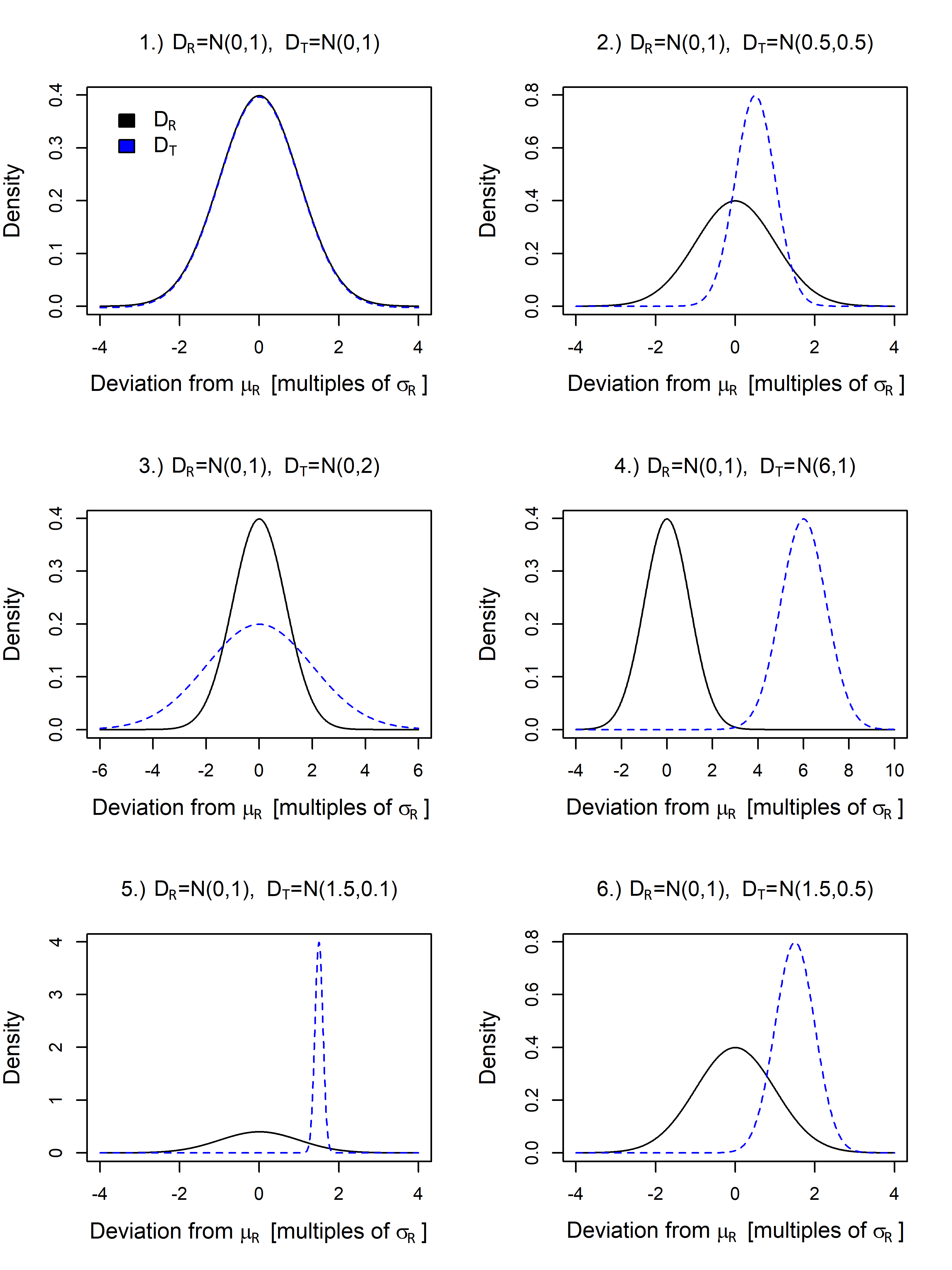}
 
\caption{Six situations one might encounter in practice. Situations 1 and 2: clear analytical similarity, situations 3 and 4: clear non-similarity, situations 5 and 6: decision depends on the extent of the covering considered necessary.} \label{eq_noneq1}
\end{figure*}

Figure~\ref{eq_noneq1} illustrates the problem at hand for several different situations in the case of normal distributions. 
When one wants to stay within the range given by $D_{R}$, one intuitively considers $D_T$ reasonably covered in scenarios 1 and 2, but rejects proper covering for scenarios 3 and 4. As demonstrated in situations 5 and 6, even in the relatively simple case of two normal distributions, it is not always obvious when to assume or reject reasonable covering. The decision depends on the, yet to be defined, \hypertarget{overlap}{\textit{degree of covering}} of the range of $D_{T}$ by the range of $D_{R}$ considered necessary. Figure~\ref{eq_noneq1} also shows that just using a combination of mean and variance comparison would allow to properly test for covering in situations 1, 3 and 4, but would be totally unsuitable for situations 2, 5 and 6, since in the latter three cases, covering would be rejected merely because of the different mean values. It obviously makes sense in the case of normal distributions to demand that the variability of the test distribution is at most that of the reference distribution. However, examples can easily be found where this need not be the case, e.g., when $D_{R}$ is normal and $D_{T}$ is a bimodal distribution. In any case, it would be too strict to demand similarity of mean values. In particular, the maximum allowed variability depends on the difference in means (and vice versa), and both are limited by the range of $D_{R}$. Note also that the concept of covering, as suggested, is not symmetric. When $D_T$ is covered by $D_R$, this does not generally mean that the reverse also holds.\\

\begin{figure*}[h!]
 \centering

 \includegraphics[width=12.5cm,  keepaspectratio,]{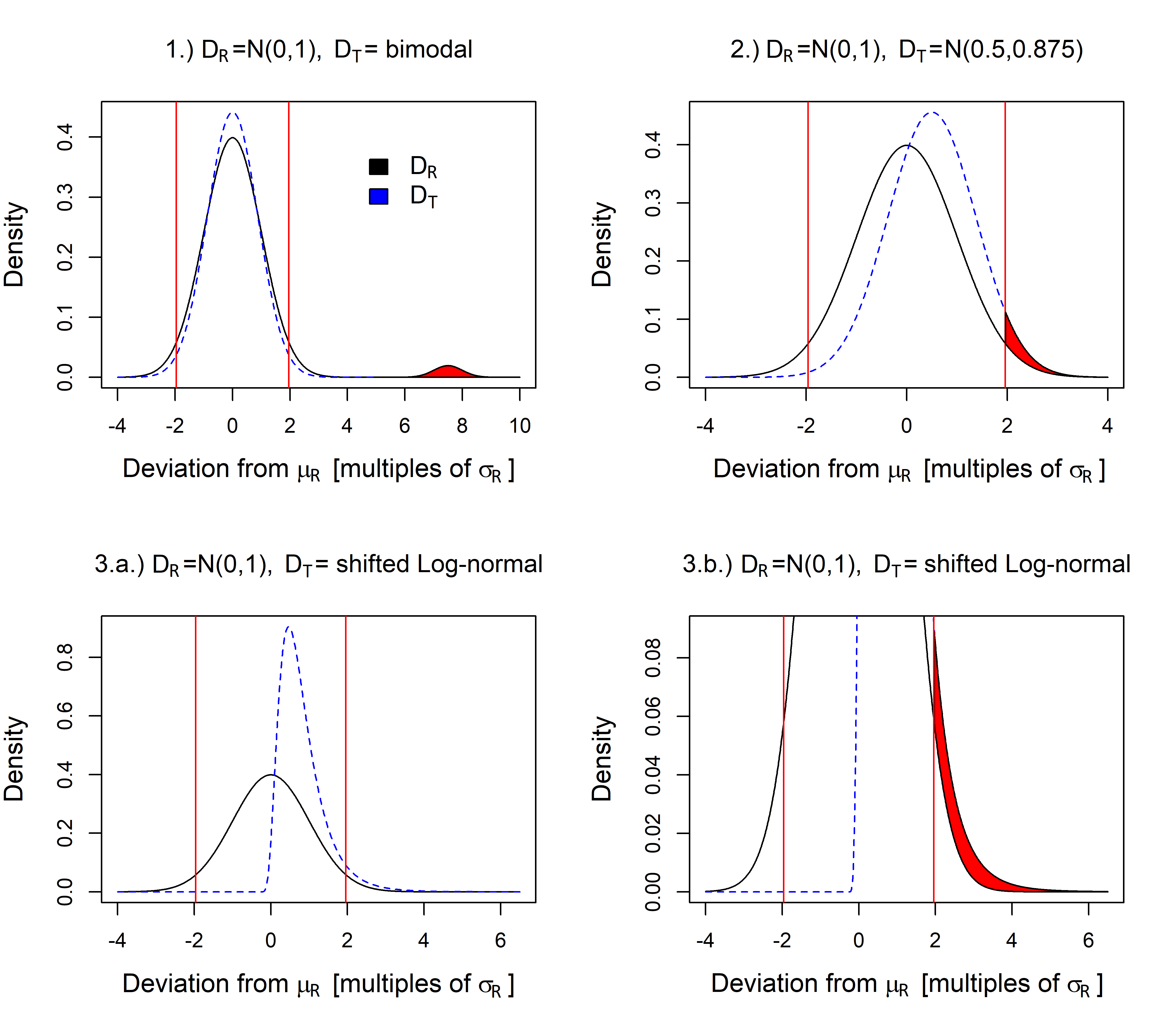}

 \caption{Increased risk of \qq{extreme} realizations of the test product compared to the reference product. In all three situations, 95\% of the probability mass of $D_{T}$ is within the range limited by the 2.5\%/97.5\% quantiles of $D_{R}$ (red vertical lines), but in all situations there is an increased risk of \qq{extreme} realizations of the test product compared to that of the reference (red area).} \label{nonnorm}
\end{figure*}

Defining what is meant by a \textit{suitable covering} of one distribution by another is of course not a trivial task. Consider, for example, the situation where a chosen fraction of the probability mass of several different distributions $D_{T}$ is within a range deduced from $D_{R}$: Assume one demands that at least 95\% of the probability mass of $D_{T}$ is within the 2.5\% and the 97.5\% quantiles of $D_{R}$. Figure~\ref{nonnorm} shows three examples where this demand is met, but the distributions still do not seem to overlap properly. In those examples, again, $D_{R} = N(\mu_{R}, \sigma_{R} )$ but now $D_{T}$ is not always normal. In situation 1 it is bimodal  and in situation 3 it is heavy tailed. Different types of distributions have been chosen to illustrate that this problem can of course occur with at least all types of distributions with non-vanishing densities on $\mathbb{R}$ or $\mathbb{R}_{\geq 0}$. Further, this issue is not always as obvious as in situation 1. In situations 2 and 3 it is more subtle, but all examples have in common that a chosen proportion of the probability mass of $D_{T}$ being within certain bounds derived from $D_{R}$ does not necessarily imply that the tails, i.e., the \qq{extreme} realizations of the total production of the test product, are within acceptable limits. Therefore, when no knowledge is available concerning specified values and allowed tolerances, one has to ensure not only that a predefined fraction of the probability mass of $D_{T}$ lies well within the range given by $D_{R}$, but also that the tails of $D_{T}$ do not lie beyond the range given by the tails of $D_{R}$.\\

In Section 2, a mathematical framework for the comparison of distributions  is introduced. Based on this framework a statistical test for this comparison in the context of normal distributions is proposed in Section 3. Section 4 presents the results of simulations and Section 5 the analysis of real world data. Section 6 contains a summary and a short outlook on the situation where additional knowledge on permissible tolerances is available. Some additional information on the calculation of generalized p-values, arguments on the asymptotic behavior of the proposed test and some additional simulation results can be found in the appendix.

\section{\emph{$\kappa$-cover}: A framework for a range-based comparison of distributions} 

The central problem of a range-based comparison of continuous distributions is to define what is meant by \qq{the range of distribution $D_{T}$ is (suitably) covered by the range of distribution $D_{R}$}. An appropriate definition of the covering of one distribution by another is obviously a necessary prerequisite for the development of suitable statistical tests. Therefore, the concept of $\boldsymbol{\kappa -cover}$ is introduced in this work which incorporates the reasoning given in Section 1. Based on this concept, $D_R$ is assumed to cover the range of $D_{T}$ to a certain extent, irrespective of the distributions at hand (i.e., form, location, variability) if certain conditions are met. It allows the degree to which one distribution is covered by another to be determined in a quantifiable manner and it simultaneously guarantees that the tails of $D_{T}$ are under control. Formally, this concept can be stated as follows:\\

\vspace*{-1\baselineskip}

\begin{definition}\label{def1}

Let $Q_{q}^{T}$ and $Q_{q}^{R}$ denote the $q$-quantiles, $q \in (0,1)$, of the distributions $D_{T}$ and $D_{R}$. For a given $\kappa \in (0,0.5], D_{R}$ constitutes a \textbf{$\boldsymbol{\kappa}$-cover} of $D_{T}$ (covers $D_{T}$ on a level of $\kappa$), $\boldsymbol{D_{R} \stackrel{\kappa}{=} D_{T}}$, iff $Q_{\beta}^{R} \leq Q_{\beta}^{T}  \ \wedge \ Q_{1-\beta}^{T} \leq Q_{1-\beta}^{R}$ \ \ $\forall \ \beta$ fulfilling $0 < \beta < \kappa.$

\end{definition} 

\vspace*{1\baselineskip}

The idea of $D_{R}$ being a $\kappa-cover$ of $D_{T}$ for a $\kappa \in (0,0.5]$ is to allow for any deviations of $D_{T}$ from $D_{R}$ within the chosen quantiles $Q_{\kappa}^{R}$ and $Q_{1-\kappa}^{R}$ as long as at least the whole probability mass of $D_{T}$ located between $Q_{\kappa}^{T}$ and $Q_{1-\kappa}^{T}$ can be found between the aforementioned quantiles of $D_{R}$, while, on the other hand, strict restrictions on the more extreme parts of $D_{T}$ (the tails, each of which contains a probability mass of $\kappa$) are demanded. This means that not only a certain central fraction of the probability mass of $D_T$ is kept under control, but that the probability of extreme values is controlled as well. Note that this relation is not symmetric, i.e., $D_{R} \stackrel{\kappa}{=} D_{T}$ does not imply $D_{T} \stackrel{\kappa}{=} D_{R}$.\\ 


Note that $D_{R} \stackrel{\kappa}{=} D_{T}$ does in general not mean, that the $\kappa$ and/or the $1-\kappa$ quantiles of the involved distributions are equal. The largest $\kappa' \in [\kappa,0.5]$ for which Def.~\ref{def1} holds true (i.e., $Q_{\kappa'}^{T} = Q_{\kappa'}^{R}      \vee    Q_{1-\kappa'}^{T} = Q_{1-\kappa'}^{R}    \iff    \nexists \kappa'' > \kappa'$ such that $D_{R} \stackrel{\kappa''}{=} D_{T}$), can be interpreted as a quantification of the actual extent to which $D_T$ is covered by $D_R$ since the larger $\kappa'$ is, the smaller the margin for the difference between the expected values of the distributions involved. Therefore, when statistically testing whether $D_{R} \stackrel{\kappa}{=} D_{T}$ holds, the choice of $\kappa$ for the corresponding statistical hypotheses is equal to the minimum extent of covering that seems acceptable, i.e., Def.~\ref{def1} allows one to define the worst case one is willing to accept. A lesser amount of deviation between $D_R$ and $D_T$ would of course also be acceptable.\\

When applying this concept to the situations depicted in Figures~\ref{eq_noneq1} \&~\ref{nonnorm} above (hereafter numbered according to the scheme: \#Figure.\#situation), one can see that in situation 1.1, $D_{R} \stackrel{\kappa}{=} D_{T}$ for $\kappa = 0.5$ and therefore $D_{R}$ is a $\kappa-cover$ of $D_{T}$ on a level of $\kappa$ for all  $0 < \kappa \leq 0.5$. In situation 1.2, $D_{R}$ is a $\kappa$-cover of $D_{T}$  for all $\kappa \leq 0.159$. The same is true for situations 1.5 and 1.6 for $\kappa \approx 0.048$ and $\kappa \approx 0.001$, respectively. In situations 1.3 and 1.4, $D_{R}$ is obviously not a $\kappa-cover$ of $D_{T}$. Whether $D_{R}$ is a $\kappa-cover$ of $D_{T}$ in situation 2.1 depends, of course, on the properties of the concrete bimodal distribution at hand. For the bimodal distribution used to illustrate situation 2.1, $D_{R}$ is actually a $\kappa-cover$ of $D_{T}$, but here $\kappa$ is not explicitly specified because it is a very small, numerically hard to compute number. In situation 2.2, $D_{R}$ is also a $\kappa-cover$ of $D_{T}$ ($\kappa \leq 0.00003167$). In situation 2.3, $D_{R}$ is not a $\kappa-cover$ of $D_{T}$ due to the heavy-tailedness of the log-normal distribution.\\

With respect to location and scale of $D_{T}$ and $D_{R}$, it follows from Def.~\ref{def1} that for a $\kappa \in (0,0.5]$ to exist such that $D_{R} \stackrel{\kappa}{=} D_{T}$ holds true,
\renewcommand{\labelenumi}{\roman{enumi})}
\begin{enumerate}
        	 \item it need not apply that $\sigma_{T} \leq \sigma_{R}$, see situation 1 in Figure~\ref{nonnorm} for an example where $\sigma_{T} > \sigma_{R}$ but still  $D_{R} \stackrel{\kappa}{=} D_{T}$ is valid for a certain (very small) $\kappa$. Nevertheless, if both distributions are normal distributions, it obviously has to apply that $\sigma_{T} \leq \sigma_{R}$ when $D_{R} \stackrel{\kappa}{=} D_{T}$.  
             \item it need not apply that $\mu_{T} = \mu_{R}$ if, depending on $\mu_{T}$ and the $\kappa$ chosen, $\sigma_{T}$ is sufficiently small. 
\end{enumerate}

The next definition connects Def.~\ref{def1} with the comparison of two physically existing entities in terms of their QAs.

\begin{definition}\label{def2}

In the case of \textbf{two entities T and R} (e.g. industrial goods), T is analytically similar to R at level $\kappa$ and with respect to a quantifiable property if for their respective distributions it holds that $D_{R}$ is a $\kappa$-cover of $D_{T}$.

\end{definition}

When applying Def.~\ref{def2}, one crucial step is of course the proper choice of $\kappa$. For example, in the development of a generic drug, one has to answer the following question: When is the analytical similarity, as quantified by $\kappa$, of the test product to the reference product great enough so as to consider this degree of analytical similarity sufficient in the context of therapeutic equivalence? For example when both distributions are normal distributions, for every difference in means a combination of $\kappa$ and $\sigma_{T}$ can be found such that $D_{R} \stackrel{\kappa}{=} D_{T}$ holds true. As stated above, the larger $\kappa$ (closer to $0.5$) for which a distribution $D_{R}$ is a $\kappa$-cover of $D_{T}$, the smaller is the maximum possible difference in mean values. Hence, when testing for analytical similarity more critical deviations require larger values of $\kappa$. A specific value for $\kappa$ that immediately comes to mind is the one which is associated with the quantiles equal to $\mu \pm 3\sigma$ in the case of a normally distributed random variable (obviously inspired by the well known $6\sigma$ approach frequently applied in the field of statistical quality control). Since the use of these quantiles, which are the $0.135\%$ and the $99.865\%$ quantile of a normal distribution, is very common in statistical applications in industry, it might often be reasonable to choose $\kappa$ equal to $0.00135$. Less extreme values such as $\kappa=0.05$ might also be considered, especially when dealing with small samples that make it more difficult to study more extreme quantiles. However, this is a question that cannot be answered by statistical reasoning alone, and needs to be answered in cooperation with pharmaceutical and health-care experts such as chemists, pharmacologists and physicians.\\

To illustrate the problems which can arise when comparing two distributions, consider again situation 1 illustrated in Figure~\ref{nonnorm}: $D_R$ and $D_T$ are continuous distributions and $Q_{\kappa}^{R} \leq Q_{\kappa}^{T}  \ \wedge \  Q_{1-\kappa}^{T} \leq Q_{1-\kappa}^{R}$ is fulfilled for $\kappa=0.025$ which guarantees that at least the central $95\%$ of the probability mass of $D_{T}$ are within the chosen limits. Nevertheless, there is no guarantee (due to the second local maximum of the density of $D_{T}$) that the restriction concerning the tails is also fulfilled, i.e., whether $Q_{q}^{T}\leq Q_{q}^{R}$ holds for all $q > 1-\kappa$. This problem can also be much less obvious in case of heavy tailed distributions, i.e., in a situation when both density functions,  $f_{T}(\cdot)$ and $f_{R}(\cdot)$, have monotonically increasing/decreasing tails.

This means that testing whether $D_{R} \stackrel{\kappa}{=} D_{T}$ is very challenging, if not impossible, when one does not know the types of distributions at hand, or at least can assume that the distributions are \qq{well-behaved} in that 


\begin{equation} Q_{\kappa}^{R} \leq Q_{\kappa}^{T} \wedge Q_{1-\kappa}^{R} \geq Q_{1-\kappa}^{T} \ \Rightarrow \ Q_{\tilde{\kappa}}^{R} \leq Q_{\tilde{\kappa}}^{T} \wedge Q_{1-\tilde{\kappa}}^{R} \geq Q_{1-\tilde{\kappa}}^{T} \ \ \forall \ 0 < \tilde{\kappa} < \kappa \tag{wb}\label{eq:wb}  \end{equation} 


If property (wb) can be assumed, it is sufficient to apply a test which, for a given $\kappa$, only deals with the $\kappa$- and $(1-\kappa)$- quantiles of the distributions to be compared. Normal distributions are well-behaved in the above-mentioned sense and also many of the approaches cited above make this assumption at least implicitly; see, e.g., \cite{Mielke.2018}, \cite{Tsong.2017} and  \cite{Boulanger.2016}. Therefore, in the further course of this work $D_{T}$ and $D_{R}$ are assumed to be normal distributions with unknown mean and variances. A further argument for assuming normally distributed QAs is that one does not want a production process to result in extreme deviations, i.e., the process will be designed in such a way that the corresponding distribution is not heavy tailed. Furthermore, due to small sample sizes, it is necessary to use a parametric testing approach since approaches which are more flexible might not have enough power. Furthermore, the assumption of normality can be tested straightforwardly with several tests suggested in literature, e.g., \cite{Jarque.1987}, \cite{Shapiro.1965} and \cite{Neyman.1937}. However, in a general framework, the normality assumption may not always be reasonable and therefore when applying the concept of $\kappa$-cover one has to be aware of the distributions one is dealing with.

\section{A statistical test for $\boldsymbol{\kappa}$-cover}
This section develops the proposed test for the property of $\kappa$-cover. The notation is introduced in \ref{Sec:notation}, while the testing problem and our null hypothesis are described in \ref{Sec:problem}. An approach for the comparison of quantile of two distribution proposed by Guo \& Krishnamoorthy\cite{Guo.2005} is reviewed in \ref{Sec:quantile_test} and we extend this test for jointly testing two quantiles. Based on this we introduce and explain our covering-test in \ref{Sec:C_test}.

\subsection{Notation}\label{Sec:notation}

\noindent $\boldsymbol{X_{T}}=(X_{T}^{1},...,X_{T}^{n_T})$ and $\boldsymbol{X_{R}}=(X_{R}^{1},...,X_{R}^{n_R})$ \label{Xx} denote random samples of size $n_T$ and $n_R$ respectively, such that $X_{T}^{i} \sim D_{T}$ and $X_{R}^{j} \sim D_{R}$, $i=1,...,n_T$ and $j=1,...,n_R$. Let $z_{\alpha}$, $\alpha \in (0,1)$, denote the $\alpha$-quantile of the standard normal distribution and, based on $\boldsymbol{X_{i}}, i=R,T$, let $\bar{X}_{i}$  and $S_{i}$ stand for the sample means and the respective sample standard deviations with realizations denoted $\bar{x}_i$ and $s_i$.\\

\noindent For normally distributed QAs, the subset \label{Theta} $\Theta \subset \mathbb{R}^4$, given as $\Theta = \{(\mu_{T},\sigma_{T},\mu_{R},\sigma_{R}) \mid \mu_{i} \in \mathbb{R}, \sigma_{i} \in \mathbb{R}_{> 0}, i \in \{R,T\} \}$, denotes the set of all admissible parameter combinations. Let \label{Tk0} $\Theta^{\kappa}_{0} \subset \Theta$, with $\kappa \in (0 , 0.5]$ arbitrary, denote the subset that consists of all parameter settings for which $D_{R}=N(\mu_{R},\sigma_{R})$ is a $\kappa$-cover of $D_{T}=N(\mu_{T},\sigma_{T})$. Thus, $\Theta^{\kappa}_{0}$ represents the parameter space of the null hypothesis. The parameter space for the alternative hypothesis is given as \label{Tk1} $\Theta^{\kappa}_{1}=\Theta \setminus \Theta^{\kappa}_{0}$, i.e., $\Theta^{\kappa}_{0}$ and $\Theta^{\kappa}_{1}$ form a partition of $\Theta$ into disjoint subsets for all $\kappa \in (0 , 0.5]$. Let \label{Tms} $\Theta_{(\mu_{R},\sigma_{R})}, \Theta^{\kappa}_{(\mu_{R},\sigma_{R}), 0}$ and $\Theta^{\kappa}_{(\mu_{R},\sigma_{R}), 1}$ be the corresponding sets for given values of $\mu_{R}$ and $\sigma_{R}$ . These sets are now subsets of $\mathbb{R}^{2}$, i.e., $\Theta^{\kappa}_{(\mu_{R},\sigma_{R}), 0}=\{(\mu_{T},\sigma_{T}) \mid \mu_{T} \in \mathbb{R}, \sigma_{T} \in \mathbb{R}_{> 0}, N(\mu_{R},\sigma_{R})$ is a $\kappa$-cover of $N(\mu_{T},\sigma_{T})\}$.\\

\subsection{The testing problem}\label{Sec:problem}

The test should meet several requirements, i.e., in particular it should allow to control the $\alpha$-error, be unbiased and have reasonable asymptotic properties. Of course, it would be desirable to have a test that need not be based on distributional assumptions, but we would expect such a test to have low power for the small sample sizes we have in mind. 
In the following, normally distributed QAs are assumed, since this is the most important distribution in practice, in particular keeping in mind that the test has to be applicable in small samples. Normal distributions fulfill (wb) and also allow to parametrically estimate the quantiles in a straightforward way. Even an unbiased estimate is possible if the appropriate correction factor is applied in the necessary estimation of the standard deviation; see \cite{Holtzman.1950}. Therefore, a test based on the comparison of quantiles will be presented for normally distributed QAs, for which, as already stated, it is sufficient to test the following hypotheses:

\begin{center}

$H_{0}: Q_{\kappa}^{R} \leq Q_{\kappa}^{T} \ \wedge \ Q_{1-\kappa}^{T} \leq Q_{1-\kappa}^{R} \vspace{-\baselineskip}$

\vspace{-.5\baselineskip}
\begin{equation}  \tag{Hyp1}\label{eq:H1} \vspace{-\baselineskip} \end{equation} 

$H_{A}: Q_{\kappa}^{R} > Q_{\kappa}^{T} \ \vee \ Q_{1-\kappa}^{T} > Q_{1-\kappa}^{R}$

\end{center}

\vspace{.5\baselineskip}

As stated by \eqref{eq:H1}, to test whether $\kappa$-cover can be assumed, the quantiles $Q_{p}^{R}$ and $Q_{p}^{T}$ have to be compared simultaneously for $p=\kappa$ and $p=1-\kappa$. This constitutes a multiple testing problem\cite{Miller.1981}, which causes inflation of the type I error. Therefore, to maintain the desired size for such a statistical test, some kind of correction is necessary, especially when the distributions are equal. This correction corresponds to that made when multiple statistical tests are performed simultaneously, but with the additional complication that the extent of the necessary correction decreases with increasing inequality of $D_R$ and $D_T$. To tackle this problem for the case of normally distributed QAs, an adaptive procedure is proposed, which guarantees the desired size and also the asymptotic unbiasedness of the resulting test. The procedure is based on the following idea: the existing generalized test statistic for the comparison of two quantiles derived by Guo \& Krishnamoorthy\cite{Guo.2005} is adjusted such that it can detect deviations with respect to two pairs of quantiles (the $\kappa$ and the $1-\kappa$ quantiles) simultaneously and by additionally accounting for error inflation due to the simultaneous comparison of more than one pair of quantiles, a test is obtained that has the above mentioned properties.\\

It is important to note here that, since the particularly important case of $D_{T}=D_{R}$ lies at the edge of the parameter set corresponding to $H_{0}$, the reversal of the hypotheses (i.e., $H_{0}: D_{R}$ is not a $\kappa$-cover of $D_{T}$) is not an option. The reversal of hypotheses is possible only if tolerance can be granted, which is, as explicitly stated above, not the situation we are dealing with in this work, since it is assumed that no knowledge about tolerances is available. A reversal of the hypotheses without additional tolerance would mean that the case of $D_{T}=D_{R}$ is part of the null hypothesis, which is to be rejected. This would have the consequence that in the case of equal distributions, similarity of products would only be accepted with a probability equal to the significance level of the test, e.g., 5\%, which would be unacceptably low.\\

\subsection{A test for jointly comparing two quantiles}\label{Sec:quantile_test}

\noindent By applying the theory of generalized p-values (see the respective work of Tsui \& Weer-ahandi\cite{Tsui.1989} and  Weerahandi\cite{Weerahandi.1995}  or the more recent literature on fiducial statistics, e.g., Hannig et al. \cite{Hannig.2016} or Murph et al. \cite{Murph.2023}, for more details) a (generalized) test statistic for the comparison of two quantiles of two distributions for the normal case can be found. This statistic is the \label{T2} $T_2$ test statistic introduced by Guo \& Krishnamoorthy\cite{Guo.2005} of which a modified version is used. The $T_2$ test statistic is (slightly) modified by replacing the two random variables in the right-hand summand of $T_2$ by $U_1$  and $U_2$ (denoted $U_R$ and $U_T$ in the $T_3$ test statistic), since the underlying quantities are identical. In the following the modified statistic is denoted \label{T3} $T_3$ as is the test based on this statistic, which is a consistent test that manages to control the chosen $\alpha$-level. Based on random samples from $D_R$ and $D_T$, $\boldsymbol{X_{R}}$ and $\boldsymbol{X_{T}}$ respectively, it is the main component of the proposed two-stage procedure.\\  

(Hyp1) consists of two pairs of quantile comparisons ($\kappa$ and $1-\kappa$) for each of which the test suggested by Guo \& Krishnamoorthy can be used. Therefore, we first review the test for each of these quantiles separately, before the adjusted version of this test for the simultaneous comparison of two pairs of quantiles is given.\\ 

For the upper tails, i.e., the part of $H_0$ of (Hyp1) based on the $1-\kappa$ quantiles, the respective hypotheses are as follows: 


\begin{equation*} \tag{Hyp2}  H_{0}: Q_{1-\kappa}^{T} \leq Q_{1-\kappa}^{R} \ \ \text{vs.} \ \ H_{A}: Q_{1-\kappa}^{T} > Q_{1-\kappa}^{R} \end{equation*} 

\vspace{.5\baselineskip}

and the $T_3$ test statistic for (Hyp2) is given by:

\vspace{-.75\baselineskip}

\begin{equation} \label{eq1} \tag{T3}
\begin{split}
T^u_{3} & = \bar{x}_{d} - \frac{\bar{X}_{d} - \mu_{d}}{(\sigma^{2}_{R}/n_{R} + \sigma^{2}_{T}/n_{T})^{0.5}} \left(\frac{\sigma^{2}_{R}}{n_{R}} \frac{s^{2}_{R}}{S^{2}_{R}} + \frac{\sigma^{2}_{T}}{n_{T}} \frac{s^{2}_{T}}{S^{2}_{T}} \right)^{0.5} + z_{1-\kappa} \left(\frac{\sigma_{R}}{S_{R}}s_{R} - \frac{\sigma_{T}}{S_{T}}s_{T} \right) \\
& = \bar{x}_{d} - Z\left(\frac{v^{2}_{R}}{n_{R}U^{2}_{R}} + \frac{v^{2}_{T}}{n_{T}U^{2}_{T}} \right)^{0.5} + z_{1-\kappa} \left(\frac{v_{R}}{\sqrt{U^{2}_{R}}} - \frac{v_{T}}{\sqrt{U^{2}_{T}}} \right)\,.
\end{split}
\end{equation}


\noindent with $\bar{x}_{d}=\bar{x}_{R}-\bar{x}_{T}$, $\mu_d = \mu_{R} - \mu_{T}$, $\bar{X}_{d}=\bar{X}_{R}-\bar{X}_{T}$ and $v^{2}_{i} = (n_{i}- 1)s^{2}_{i}, i = R, T$. Furthermore, $Z$ and $\ U^{2}_{i}, i = R, T$, are independent random variables with $Z \sim N(0,1)$, $U^{2}_{i} \sim \chi^2_{n_{i}- 1}$ and $n_i$ being the respective sample sizes.


Similarly, to test the $\kappa$-quantiles of (Hyp1), one uses \label{T3l}
\[
T^l_3=-\bar{x}_d-Z\left(\frac{v^2_R}{n_r U^2_R}+\frac{v^2_T}{n_T U^2_T}\right)^{0.5}+z_{1-\kappa}\left(\frac{v_R}{\sqrt{U^2_R}}-\frac{v_T}{\sqrt{U^2_T}}\right).
\]

Since, as already explained above, it is necessary to test both sub-hypotheses of $H_0$ of (Hyp1) simultaneously in order to test the assumption of $\kappa$-cover, a ``two-sided'' version of the $T_3$ statistic is given by

\vspace{-1\baselineskip}

\begin{equation*} \label{eq2} 
T_{|3|}=-|\bar{x}_d|-Z\left(\frac{v^2_R}{n_r U^2_R}+\frac{v^2_T}{n_T U^2_T}\right)^{0.5}+z_{1-\kappa}\left(\frac{v_R}{\sqrt{U^2_R}}-\frac{v_T}{\sqrt{U^2_T}}\right).
\end{equation*}

\vspace{.5\baselineskip}

It combines $T_3^u$ and $T_3^l$ in that it holds that $P(T_{|3|}>0)=min(P(T^u_3>0), P(T^l_3>0))$, which can easily be checked. 

Appendix \ref{App 1} contains further information on the calculation of the corresponding p-values of $T^u_3$, $T^l_3$ and $T_{|3|}$ which is based on the respective information given in Guo \& Krishnamoorthy\cite{Guo.2005} since everything stated in \cite{Guo.2005} for $T_2$ also holds for $T^u_3$ and $T^l_3$.\\

\subsection{The covering-test (C-test): An adaptive procedure based on generalized p-values}\label{Sec:C_test}

\begin{figure*}[b!]
 \centering
 \includegraphics[width=6cm,  height = 6cm]{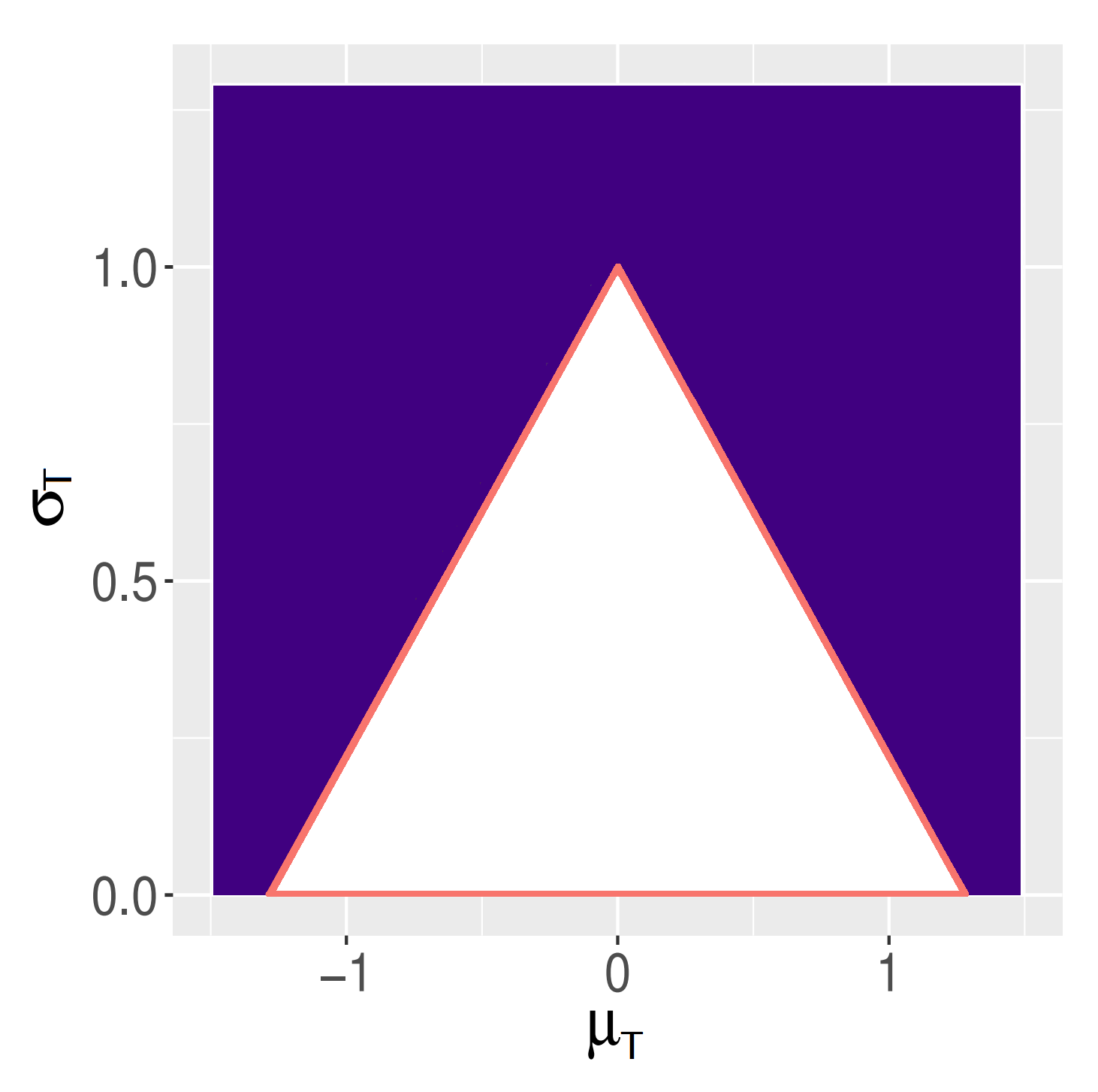}
 \caption{$D_R = N(0,1)$ and $\kappa=0.1$. Since $D_R$ is fixed, the $H_0$ parameter space is a subset of $\mathbb{R}^{2}$, otherwise it is a subset of $\mathbb{R}^{4}$. Red triangle: parameter set of the null hypothesis $\Theta^{0.1}_{(0,1),0}$ together with its boundary $\partial\Theta^{0.1}_{(0,1),0}$. White interior: $\Theta^{0.1 \mathrm{o}}_{(0,1),0}$. $\Theta^{0.1}_{(0,1),0}$ does not include the base of the red triangle.} \label{area biasedness4}
\end{figure*}

In order to get a clear picture of the problem, let's assume, w.l.o.g., that $D_R = N(0,1)$ and $\kappa \in (0,0.5]$. The corresponding parameter space of $H_0$, $\Theta^{\kappa}_{(0,1),0}$, is, depending on $\kappa$, given by a triangle like the one shown in Figure~\ref{area biasedness4}. The white interior shows \label{oms} $\Theta^{\kappa\mathrm{o}}_{(0,1),0}$ and the red isosceles triangle \label{pms} $\partial\Theta^{\kappa}_{(0,1),0}$, but, since $\sigma_T > 0$ is demanded, only the triangle without base is part of $\Theta^{\kappa}_{(0,1),0}$. Since the border of $\Theta^{\kappa}_ 0$ is of special interest with respect to the behavior of the test, it is expressly pointed out that moving along the border, i.e., the legs of the red triangle in Figure~\ref{area biasedness4}, away from equality of distributions, not only means to increase the distance between $\mu_T$ and $\mu_R$, but also to decrease $\sigma_T$ accordingly such that either $\ Q_{\kappa}^{R} = Q_{\kappa}^{T}$ or $\ Q_{1-\kappa}^{R} = Q_{1-\kappa}^{T}$ (which means that $H_0$ is still valid).\\ 


Defining the statistics $T^u_3$, $T^l_3$ and $T_{|3|}$ sets the stage for constructing a test for testing the hypotheses in (Hyp1), since the above mentioned multiple testing issue (two times two quantiles have to be compared) still awaits a solution. Just using the test statistic $T_{|3|}$, or combining $T^u_3$ and $T^l_3$ to simultaneously compare the lower and upper quantiles does not provide a satisfactory solution to the problem of constructing a test for (Hyp1). For example, when using $T_{|3|}$ and choosing \label{phi} $\kappa = \Phi(-3)$ and a significance level $\alpha$ of 0.1, the rejection probability in the case $D_T = D_R$ is significantly greater than the value it should have ($\approx 0.16$ instead of $0.1$), while the rejection probability in the case of $D_T = N(2.7 , 0.1)$ is still around $0.1$. This is due to the fact that for $D_T = N(2.7 , 0.1)$ effectively only the one-sided test applies because only one of the quantiles will violate $H_0$ as the other quantile of $D_T$ does not lie on the boundary of $H_0$. 
Combining two tests and choosing the significance level of both tests such that the resulting significance level of the combined tests is equal to $\alpha$ for $D_T=D_R$ does also not work properly. The resulting test will reject $H_0$ less likely along the border of \label{Tk0g} $\Theta^{\kappa}_0$, $\partial\Theta^{\kappa}_0$, i.e., when $\mu_T$ is closer to $Q_{\kappa}^{R}$ or $Q_{1-\kappa}^{R}$ respectively.\\ 

Therefore, when simultaneously comparing two sets of quantiles (here the $\kappa$ and the $1-\kappa$ quantiles of $D_R$ and $D_T$), given $H_0$ holds true, a need for some kind of correction (similar to Bonferroni correction) arises when in both sets of quantiles the respective quantiles do not differ much. This is, of course, especially true for equal distributions, but the resulting need for correction also persists within a certain vicinity of $D_T = D_R$, but decreases with increasing inequality of $D_T$ and $D_R$. This means when moving away from the case of equal distributions along the border of the $H_0$ parameter space, the larger $|\mu_d|$ gets, the less correction is needed. Therefore, it would be necessary to detect, whether correction is needed and to what extent it has to be applied.\\

To address the problems arising from the multiple testing problem, a test procedure termed $covering$- or $C$-test is proposed. It combines the test statistic for comparing the quantiles of interest of $D_R$ and $D_T$ with respect to \eqref{eq:H1}, with an implicit pretest intended to decide, whether the adjusted version of the $T_3$ test, $T_{|3|}$, has to be used for comparing these quantiles simultaneously or it is sufficient to just compare one of these quantiles based on the unadjusted statistic. The $C$-test can be interpreted as an adaptive procedure consisting of a pretest (step 1) and the actual comparison of the quantiles of interest (step 2), but both steps are performed simultaneously and are suitably combined into a single test statistic. The resulting test statistic is the test statistic of the C-test, which is constructed such that an overall significance level of \label{alpha} $\alpha$ is maintained.\\

When both distributions are equal, the necessary adjustment of the p-value computed with respect to $T_{|3|}$ to ensure that the test size equals $\alpha$ is achieved by computing the resulting p-value as 
\[
p_{|3|}\cdot K, 
\]
where $K > 1$ is a constant whose determination is analytically challenging since the tests involved are not independent of each other. This adjustment of the p-value is due to the fact that given that $D_R=D_T$ either only one of the sub-hypothesis in $H_0$ concerning the $\kappa$ or the $1-\kappa$ quantile might be rejected or both might be rejected simultaneously. Therefore, it is determined using simulations for given sample sizes, $\kappa$ and the desired significance level $\alpha$.\\

The adjustment of the p-value is only necessary to control the size when $D_R=D_T$ or when $D_R\neq D_T$ but the distributions are still close to each other in the sense that $H_0$ holds and both tests, i.e., $T_3^u$ and $T_3^l$, still reject with positive probability such that the combined rejection probability exceeds the chosen significance level $\alpha$. If, in the latter case either the $\kappa$ or the $1-\kappa$ quantiles are equal, $T_{|3|}$ converges to either $T^u_3$ or $T_3^l$ with increasing sample sizes and increasing mean value difference $|\mu_d|$, i.e., the need for the adjustment used in the computation of $p_{|3|}$ vanishes and is no longer needed if the sample sizes with respect to $|\mu_d|$ are large enough.\\

A priori, one does not know which situation applies, i.e., whether both quantiles have to be tested or only one of them. Therefore, an adaptive test statistic is suggested that is based on the following idea: Perform a pre-test of the hypotheses
\[
H_{0l}: Q_{\kappa}^R \geq Q_{\kappa}^T \text{ vs. } H_{1l}: Q_{\kappa}^R < Q_{\kappa}^T
\]

\vspace*{-.7\baselineskip} 

and
\[
H_{0u}: Q_{1-\kappa}^R \leq Q_{1-\kappa}^T \text{ vs. } H_{1u}: Q_{1-\kappa}^R > Q_{1-\kappa}^T,
\]


hence reversing the original null and alternative. These can be tested using $T_3^l$ and $T_3^u$, but computing the p-values as \label{bpul} $\bar{p}_{x}= P(T_3^{x}<0), x=u,l$. If either of these hypotheses is rejected one can be fairly certain that the distributions are not equal and that it is sufficient to test either the upper or the lower quantile. In that case the p-value does not need to be adjusted by \label{Kadj} $K$ to maintain the correct size of the test. To determine whether a one-sided  or a \qq{two-sided} test  has to be performed, \label{bpmin} $\bar{p}_{min}:=\min(\bar{p}_l,\bar{p}_u)$ needs to be compared to a predefined threshold/tuning parameter  \label{alp} $\alpha_p \in (0,1)$. Let \label{ind} $I(\cdot)$ be the indicator function. Then, the p-value of the resulting test (the C-test) is given by
\begin{equation}\label{eq:p_adj}\tag{$p_{C}$}
p_C=I(\bar{p}_{min}<\alpha_p)\cdot p_{|3|}+(1-I(\bar{p}_{min}<\alpha_p))\cdot p_{|3|}\cdot K
\end{equation}

This is easy to implement and numerically one can see that for a properly determined $K$ and reasonable values of $\alpha_p$ (e.g., $\alpha_p \in [0.01,0.1]$) the test has the desired size $\alpha$, while maintaining good power. In Appendix \ref{App 2} we sketch arguments for the asymptotic validity of the test.\\

Note that in finite samples we may have the situation that $D_T\neq D_R$ but we still use the \qq{two-sided} (i.e. size-corrected) part of the test with positive probability. It is non-trivial that the test still has rejection probability $\leq \alpha$ due to the dependence of all terms involved in \eqref{eq:p_adj} and the conditional nature of the components of the formula. However, our simulations in the next section demonstrate that this is indeed that case.\\

\section{Monte Carlo simulations}
In this section the results of a number of Monte Carlo simulations to study the finite sample behavior of the proposed C-test are presented. Recall that normally distributed data is assumed for the quality attributes. Without loss of generality it is further assumed the distribution of the reference product, $D_R$, to be $N(0,1)$. The parameters of the distribution of the test product, $D_T$, are denoted by $\mu_T$ and $\sigma_T$, and are varied depending on whether the behavior under $H_0$ or under $H_1$ is studied. The sample size is the same for both products, and $n=10, 25, 50, 100$ is considered, reflecting situations when it is rather costly to obtain measurements of quality attributes for reference and test products and also situations where plenty of data is available. The $\kappa$ values $0.1$, $0.05$ and $\Phi(-3)=0.00135$ are selected for the quantiles of interest. The tuning parameter of the test $\alpha_p$, interpreted as the significance level of the pre-test, takes on the values $0.01$, $0.05$, $0.1$ and $0.2$. To evaluate the generalized p-values of each test $N=10,000$ random numbers are drawn. Unless stated otherwise, the results are based on $M=100,000$ Monte Carlo simulations. Finally, all tests have a nominal size of $\alpha=0.05$.

Section \ref{Sec:K} is concerned with the determination of the adjustment parameter $K$ and determined values are tabulated for several different parameter combinations, while Sections \ref{Sec:size} and \ref{Sec:power} respectively show the results for the size and power of the test.

\subsection{Determining the adjustment parameter $K$}\label{Sec:K}

\begin{table}[b!]
\caption{\label{Kor} Simulated values for the adjustment constant $K$ for several different combinations of  $\kappa, n$ and $\alpha_p$.}
\centering

\setlength\extrarowheight{-3pt}

\begin{tabular}{rrrrrr}

  \hline
$\kappa$ & $n$ & $\alpha_p=0.01$ & $\alpha_p=0.05$ & $\alpha_p=0.1$ & $\alpha_p=0.2$ \\ 
  \hline
0.1 & 10 & 1.81 & 1.88 & 1.99 & 2.31 \\ 
  0.1 & 25 & 1.94 & 2.04 & 2.21 & 2.72 \\ 
  0.1 & 50 & 1.97 & 2.08 & 2.26 & 2.83 \\ 
  0.1 & 100 & 1.99 & 2.12 & 2.31 & 2.93 \\ 
\hline
0.05 & 10 & 1.80 & 1.83 & 1.87 & 2.00 \\ 
  0.05 & 25 & 1.89 & 1.93 & 1.99 & 2.16 \\ 
  0.05 & 50 & 1.92 & 1.96 & 2.03 & 2.22 \\ 
  0.05 & 100 & 1.94 & 1.99 & 2.05 & 2.25 \\ 
\hline 
  0.00135 & 10 & 1.61 & 1.61 & 1.61 & 1.62 \\ 
  0.00135 & 25 & 1.67 & 1.67 & 1.67 & 1.68 \\ 
  0.00135 & 50 & 1.69 & 1.69 & 1.69 & 1.70 \\ 
  0.00135 & 100 & 1.70 & 1.70 & 1.70 & 1.70 \\ 
   \hline
\end{tabular}
\end{table}

Let $\theta_e \in \Theta$ represent the parameter combination in the case of equal distributions, i.e., when $D_R=D_T$. Recall that the adjustment parameter $K$ needs to be determined numerically in order to ensure correct size of the test and that asymptotically its choice is only relevant for $\theta_e$ due to the consistency of the pretest. However, for finite sample sizes, the rejection probability for parameter combinations within a certain vicinity of $\theta_e$ is smaller than the desired size of the test. In addition, due to the continuity of the test statistic, this undesired behavior is not only present on the border of $\Theta_0^{\kappa}$, but also occurs within an adjacent area of $\Theta_1^{\kappa}$, biasing the test. This behavior is due to the following two effects which determine the rejection probability of the test on $\partial\Theta^{\kappa}_{0}$:

\renewcommand{\labelenumi}{\roman{enumi})}
\begin{enumerate}
      \item Since one of the pre-test null hypotheses is violated when moving away from $\theta_e$ along $\partial\Theta^{\kappa}_{0}$, not only the probability that both pre-tests are not significant decreases, but also the probability that $p_{|3|} < 0.05$  when both pretests are not significant. 
      \item The probability for rejection in combination with one significant pretest is increasing.
\end{enumerate}

These insights are based on careful inspection of simulations for the terms entering the test statistic. The first effect would make it necessary to reduce $K$ accordingly in order to compensate for this behavior, which is not possible as this would require knowledge of the underlying distributions. Even the second effect cannot initially compensate for the loss of rejection probability caused by the fact that the corrected  $T_{|3|}$ test operates apart from $\theta_e$, but, as the distance increases, the first effect becomes less and less pronounced, while the second effect becomes more and more pronounced, until finally the distance is so large that, due to the consistency of the test, all rejections come from the one sided test. The extent of the vicinity of $\theta_e$ in which this phenomenon of decreased rejection probability can be observed decreases with increasing sample sizes. Figure \ref{Sim1} in Appendix \ref{App 3} shows an illustration of this behavior for several sample sizes on $\partial\Theta^{\kappa}_{(0,1)0}$ for $\kappa=0.05$.\\ 

$K$ is determined using $1,000,000$ Monte Carlo simulations assuming $D_R=D_T=N(0,1)$ and all other parameters as stated above. Table \ref{Kor} shows that $K$ varies with the chosen quantile $\kappa$, the sample size $n$ and the parameter $\alpha_p$. Interestingly, for $\kappa=0.00135$ it is almost constant over different values of $\alpha_p$.

\subsection{Size}\label{Sec:size} 
For studying the size of the C-test different parameter combinations from $\partial\Theta^{\kappa}_{(0,1),0}$, the boundary of $H_0$, are selected. To be specific, $\mu_T=0,0.5, 1$ and $\sigma_T=(z_{1-\kappa} - \mu_T)/z_{1-\kappa}$ are chosen, ensuring that the parameter vector lies on the boundary of $H_0$. The case $\mu_T > 0$ implies that the $1-\kappa$ quantiles of $D_R$ and $D_T$ are equal, while the $\kappa$ quantile is smaller for $D_R$. Table \ref{tab:size} shows the results. Not surprisingly the test is correctly sized for $\mu_T=0$ due to the fact that the adjustment parameter $K$ was selected to ensure proper size whenever $D_R=D_T$. When $\mu_T=1$ the test also shows proper rejection probability for all sample sizes and $\kappa$ values with the exception of $\kappa=0.00135$ in combination with small sample sizes, where the test is a bit undersized. When $\mu=0.5$ the size distortions are more prominent and the test is a bit undersized for small sample sizes and for smaller values of $\kappa$. It is notable that the size is closer to the nominal size for larger values of $\alpha_p$. Hence, it is concluded that the test is a bit undersized whenever $D_R$ and $D_T$ are similar but not equal. This corresponds to situations close to the tip of the triangle representing $\partial\Theta^{\kappa}_{(0,1),0}$ depicted in Figure \ref{area biasedness4}.

To study these size distortions more carefully extensive simulations have been performed; see Appendix \ref{App 3} for some of these results.

\begin{table}[t!]
\caption{\label{tab:size} Size of the C-Test}

\setlength\extrarowheight{-3pt}

\centering
\begin{tabular}{rrrrrrr}
  \hline

$\mu_T$  & $\kappa$ & $n$ & $\alpha_p=0.01$ & $\alpha_p=0.05$ & $\alpha_p=0.1$ & $\alpha_p=0.2$ \\ 
  \hline
0.0 & 0.10000 & 10 & 0.050 & 0.049 & 0.049 & 0.050 \\ 
  0.0 & 0.10000 & 25 & 0.050 & 0.050 & 0.050 & 0.050 \\ 
  0.0 & 0.10000 & 50 & 0.049 & 0.049 & 0.049 & 0.049 \\ 
  0.0 & 0.10000 & 100 & 0.051 & 0.051 & 0.051 & 0.051 \\ 
\hline
  0.0 & 0.05000 & 10 & 0.051 & 0.051 & 0.051 & 0.051 \\ 
  0.0 & 0.05000 & 25 & 0.049 & 0.049 & 0.049 & 0.050 \\ 
  0.0 & 0.05000 & 50 & 0.050 & 0.050 & 0.050 & 0.050 \\ 
  0.0 & 0.05000 & 100 & 0.050 & 0.050 & 0.050 & 0.050 \\ 
\hline 
  0.0 & 0.00135 & 10 & 0.050 & 0.050 & 0.050 & 0.050 \\ 
  0.0 & 0.00135 & 25 & 0.049 & 0.049 & 0.049 & 0.049 \\ 
  0.0 & 0.00135 & 50 & 0.051 & 0.051 & 0.051 & 0.051 \\ 
  0.0 & 0.00135 & 100 & 0.049 & 0.049 & 0.049 & 0.049 \\ 
\hline 
  0.5 & 0.10000 & 10 & 0.035 & 0.042 & 0.045 & 0.048 \\ 
  0.5 & 0.10000 & 25 & 0.045 & 0.048 & 0.049 & 0.049 \\ 
  0.5 & 0.10000 & 50 & 0.051 & 0.051 & 0.051 & 0.051 \\ 
  0.5 & 0.10000 & 100 & 0.051 & 0.051 & 0.051 & 0.051 \\ 
\hline
  0.5 & 0.05000 & 10 & 0.031 & 0.036 & 0.039 & 0.043 \\ 
  0.5 & 0.05000 & 25 & 0.037 & 0.043 & 0.046 & 0.047 \\ 
  0.5 & 0.05000 & 50 & 0.048 & 0.050 & 0.050 & 0.051 \\ 
  0.5 & 0.05000 & 100 & 0.051 & 0.051 & 0.051 & 0.051 \\ 
\hline 
  0.5 & 0.00135 & 10 & 0.032 & 0.032 & 0.032 & 0.034 \\ 
  0.5 & 0.00135 & 25 & 0.031 & 0.032 & 0.033 & 0.037 \\ 
  0.5 & 0.00135 & 50 & 0.032 & 0.036 & 0.039 & 0.044 \\ 
  0.5 & 0.00135 & 100 & 0.037 & 0.044 & 0.047 & 0.049 \\ 
\hline 
  1.0 & 0.10000 & 10 & 0.052 & 0.052 & 0.052 & 0.052 \\ 
  1.0 & 0.10000 & 25 & 0.051 & 0.051 & 0.051 & 0.051 \\ 
  1.0 & 0.10000 & 50 & 0.051 & 0.051 & 0.051 & 0.051 \\ 
  1.0 & 0.10000 & 100 & 0.051 & 0.051 & 0.051 & 0.051 \\ 
\hline 
  1.0 & 0.05000 & 10 & 0.050 & 0.051 & 0.052 & 0.052 \\ 
  1.0 & 0.05000 & 25 & 0.052 & 0.052 & 0.052 & 0.052 \\ 
  1.0 & 0.05000 & 50 & 0.051 & 0.051 & 0.051 & 0.051 \\ 
  1.0 & 0.05000 & 100 & 0.051 & 0.051 & 0.051 & 0.051 \\ 
\hline 
  1.0 & 0.00135 & 10 & 0.034 & 0.037 & 0.040 & 0.044 \\ 
  1.0 & 0.00135 & 25 & 0.042 & 0.049 & 0.051 & 0.052 \\ 
  1.0 & 0.00135 & 50 & 0.051 & 0.052 & 0.052 & 0.052 \\ 
  1.0 & 0.00135 & 100 & 0.050 & 0.050 & 0.050 & 0.050 \\ 
   \hline
\end{tabular}
\end{table}


\subsection{Power}\label{Sec:power}

\begin{table}[t!]
\caption{\label{tab:power} Power of the C-test}

\setlength\extrarowheight{-3pt}

\centering
\begin{tabular}{rrrrrrrr}
  \hline
  $\mu_T$ & $\sigma_T$ & $\kappa$ & $n$ & $\alpha_p=0.01$ & $\alpha_p=0.05$ & $\alpha_p=0.1$ & $\alpha_p=0.2$ \\ 
  \hline
0.0 & 1.2 & 0.10000 & 10 & 0.106 & 0.104 & 0.102 & 0.098 \\ 
  0.0 & 1.2 & 0.10000 & 25 & 0.165 & 0.161 & 0.155 & 0.140 \\ 
  0.0 & 1.2 & 0.10000 & 50 & 0.253 & 0.246 & 0.235 & 0.212 \\ 
  0.0 & 1.2 & 0.10000 & 100 & 0.413 & 0.400 & 0.385 & 0.345 \\ 
\hline 
0.0 & 1.2 & 0.05000 & 10 & 0.111 & 0.110 & 0.110 & 0.106 \\ 
  0.0 & 1.2 & 0.05000 & 25 & 0.183 & 0.181 & 0.179 & 0.173 \\ 
  0.0 & 1.2 & 0.05000 & 50 & 0.287 & 0.284 & 0.279 & 0.267 \\ 
  0.0 & 1.2 & 0.05000 & 100 & 0.468 & 0.464 & 0.458 & 0.442 \\ 
\hline 
  0.0 & 1.2 & 0.00135 & 10 & 0.121 & 0.121 & 0.121 & 0.121 \\ 
  0.0 & 1.2 & 0.00135 & 25 & 0.207 & 0.207 & 0.207 & 0.206 \\ 
  0.0 & 1.2 & 0.00135 & 50 & 0.331 & 0.331 & 0.331 & 0.331 \\ 
  0.0 & 1.2 & 0.00135 & 100 & 0.537 & 0.537 & 0.537 & 0.537 \\ 
\hline
  0.5 & 1.0 & 0.10000 & 10 & 0.115 & 0.121 & 0.127 & 0.134 \\ 
  0.5 & 1.0 & 0.10000 & 25 & 0.250 & 0.270 & 0.282 & 0.296 \\ 
  0.5 & 1.0 & 0.10000 & 50 & 0.475 & 0.505 & 0.520 & 0.535 \\ 
  0.5 & 1.0 & 0.10000 & 100 & 0.785 & 0.805 & 0.813 & 0.819 \\ 
\hline 
  0.5 & 1.0 & 0.05000 & 10 & 0.098 & 0.101 & 0.105 & 0.111 \\ 
  0.5 & 1.0 & 0.05000 & 25 & 0.202 & 0.216 & 0.228 & 0.242 \\ 
  0.5 & 1.0 & 0.05000 & 50 & 0.381 & 0.410 & 0.426 & 0.444 \\ 
  0.5 & 1.0 & 0.05000 & 100 & 0.679 & 0.707 & 0.718 & 0.726 \\ 
\hline 
  0.5 & 1.0 & 0.00135 & 10 & 0.074 & 0.074 & 0.074 & 0.075 \\ 
  0.5 & 1.0 & 0.00135 & 25 & 0.121 & 0.121 & 0.123 & 0.128 \\ 
  0.5 & 1.0 & 0.00135 & 50 & 0.195 & 0.200 & 0.208 & 0.222 \\ 
  0.5 & 1.0 & 0.00135 & 100 & 0.352 & 0.374 & 0.392 & 0.412 \\ 
   \hline
\end{tabular}
\end{table}

\vspace*{1\baselineskip} 

To analyze the power of the test, again a standard normal distribution for $D_R$ and two alternative distributions for $D_T$ are considered. The first case is a ``symmetric/two-sided'' deviation from $H_0$ that may be visualized by moving up from the tip of the triangle in Figure \ref{area biasedness4} that represents the parameter combinations under the null. Specifically, $\mu_T=0$ and $\sigma_T=1.2$ are considered implying that the $\kappa$ and $1-\kappa$ quantiles of $D_T$ have the same distance from the respective quantiles of $D_R$. The second alternative considers the ``asymmetric/one-sided'' case where only the $1-\kappa$ quantile of $D_T$ violates $H_0$ using $\mu_T=0.5$ and $\sigma_T=1$. The results shown in Table \ref{tab:power} allow for a few conclusions. Naturally the power increases with the sample size. For $n=10$ and $n=25$ the test has low power. The power is also larger for more central quantiles, i.e. for larger values of $\kappa$, which is to be expected given the sparsity of observations in the tails for more extreme quantiles. Most interesting is the choice of the parameter $\alpha_p$, which can be chosen by the researcher. In the symmetric case the test has more power for smaller values of $\alpha_p$, although this effect becomes less pronounced for smaller $\kappa$ and completely disappears for $\kappa=0.00135$. In contrast, the opposite is observed in the second (asymmetric) alternative, where the power increases in $\alpha_p$. In this instance this effect does not depend one the choice of $\kappa$. This phenomenon is due to the following effects:

\renewcommand{\labelenumi}{\roman{enumi})}
\begin{enumerate}
        	 \item The $T_{|3|}$ test statistic is sensitive to violations with respect to both quantiles and has therefore more power than the test statistic which focuses only on one of the quantiles when both subhypotheses of (Hyp1) are violated.
             \item The opposite is true when only one of the subhypotheses of (Hyp1) is violated, i.e., the test statistic which focuses only on one of the quantiles has more power in the latter situation.
\end{enumerate}

Thus, the larger $\alpha_p$, the less likely the $T_{|3|}$ test statistic goes into action and vice versa. Therefore, larger values for $\alpha_p$ reduce the power of the test when both subhypotheses are violated and increase the power of the test when only one of theses hypotheses is violated.\\

In conclusion, larger values of $\alpha_p$ are recommended, particularly, when one is interested in rather small values of $\kappa$, since the potential gain in power in the asymmetric case seems to outweigh the possible loss in power in the symmetric case, and one does not know a priori which case applies.\\

In addition to the simulation results given in the tables \ref{Kor} - \ref{tab:power}, a lot more information regarding the simulation results is given by Figures \ref{Sim1} and \ref{fig:contour}, see Appendix \ref{App 3} for these figures and additional information regarding the results.

\section{Application}

In this section we demonstrate the application of the proposed C-test to a real world dataset from the pharmaceutical industry, which was generated in the development of a generic drug. The data contains measured concentrations of the active pharmaceutical ingredient (API) of the test and the reference product. For the API, there is an upper as well as a lower bound to its concentration. Unfortunately, the specific limits are not known to the manufacturer of the generic drug, who must now prove to the authorities that the API concentration of his product is within a reasonable range. In addition, there is also no information on permissible tolerances. Therefore the manufacturer has to compare the API concentration range of his product to the range of the originator. Due to the variability of the API concentration present in both products, approaches like the plain comparison of mean values or the testing of equality of distributions (e.g. by using a Kolmogorov-Smirnov-test) must be considered unsuitable. The min/max-approach can also not be recommended since it is an overly conservative method. For example, \cite{Niazi.2022} show a rejection rate of 75\% in the case of equal distributions. Due to different reasons several other existing approaches for range comparisons also do not appear to be applicable. These issues are, in the first place, a lack of information on tolerances, but also, as discussed for example in the EMA reflection paper\cite{EMA.2021}, other problems, such as fundamental statistical concerns such as those that arise when applying tolerance intervals.\\  

\begin{figure*}[b!]
 \centering
 \includegraphics[width=16cm]{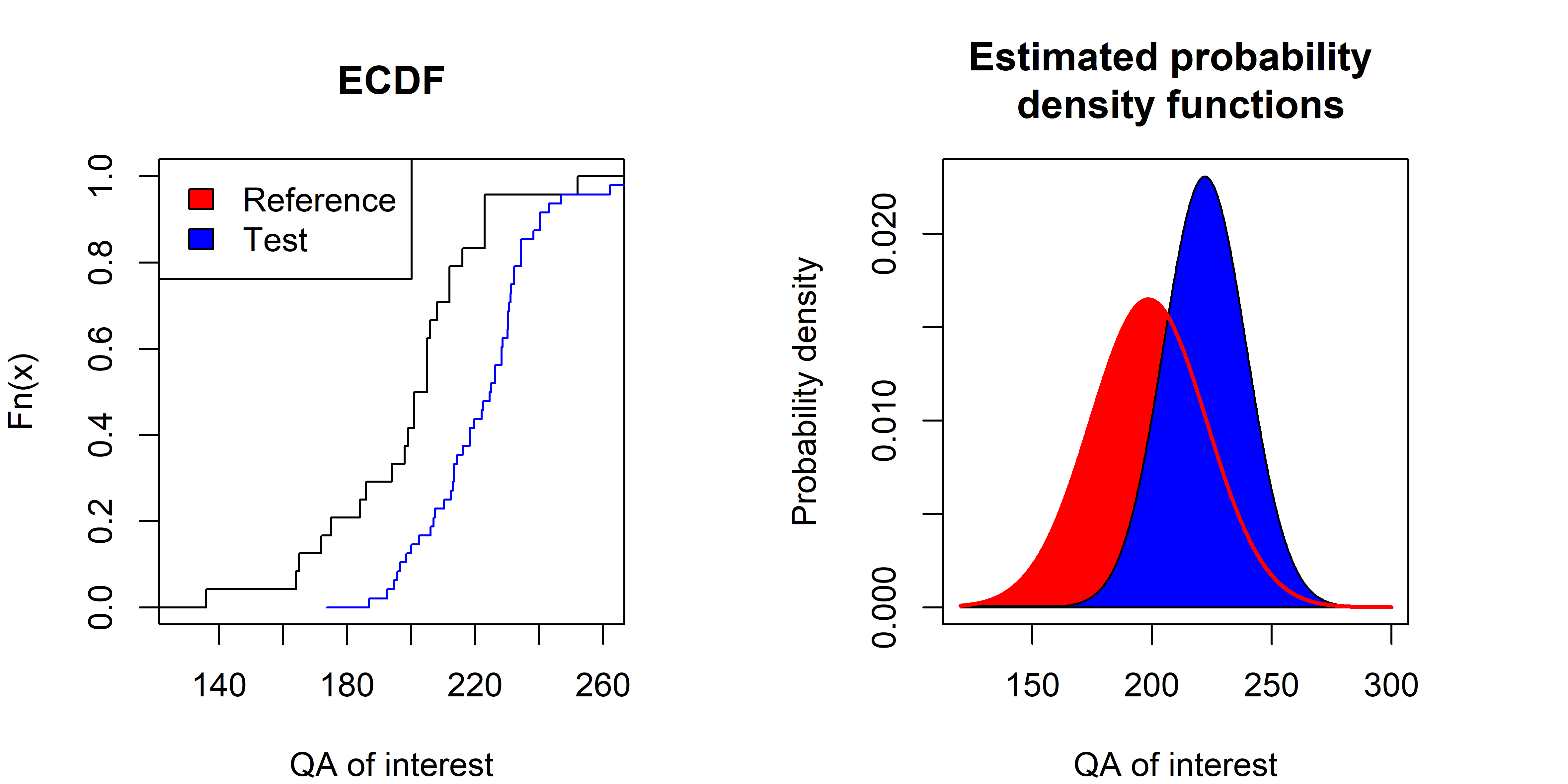}
 \caption{Empirical distribution function and estimated density function (the latter is based on the assumption of normally distributed observations) for
the data from industry with respect to reference and test product.} \label{indat}
\end{figure*}

The test product was classified as borderline with regard to the QA under discussion, but was eventually assessed as acceptable by the responsible subject matter experts. Therefore we consider this data set to be particularly suitable for illustrating the problem at hand. The empirical distribution functions and the respective normal distributions with respect to reference and test product as deduced from the data is illustrated graphically in Fig.~\ref{indat} (the assumption of normally distributed data was not rejected by the Shapiro-Wilk-test). More information on the product and the analyzed substance cannot be given due to a non-disclosure agreement between the authors and the producer of the generic drug. As can be seen, the estimated distribution of the QA of the test product appears to be covered by the estimated distribution of the QA of the reference product with respect to a properly chosen $\kappa$. Due to the reasonable number of observations (24 for the reference and 48 for the test product), the $C$-test should have acceptable power and applying a properly parameterized $C$-test should provide a good compromise here between consumer and producer risk. As already discussed, the application of the $C$-test require schoosing two parameters, namely the tuning parameter $\alpha_p$ and the quantile defining $\kappa$. Especially the latter is essential for the hypotheses one wants to consider since the minimum required extent to which $D_T$ should be covered by $D_R$ is, as discussed above, determined by $\kappa$.\\

Fig.~\ref{indat} shows, that $D_T$ deviates greatly from $D_R$ with respect to quantiles below the third quartile, which has the consequence that not only the left tail but also the central fraction of the probability mass of $D_T$ is significantly shifted away from the corresponding regions of $D_R$. Nevertheless, since $\sigma_T < \sigma_R$, there exists a $\kappa \in (0,0.5]$ such that $D_{R} \stackrel{\kappa}{=} D_{T}$ holds true. Due to the deviations in the central part of $D_R$ (the left tail of $D_T$ is well covered by the left tail of $D_R$ and therefore no reason of concern), the largest $\kappa \in (0,0.5]$ for which $D_{R} \stackrel{\kappa}{=} D_{T}$ holds true is therefore considerably smaller than 0.5, or, more formally, $\kappa':=\textrm{max}\{\kappa \in (0,0.5] | D_{R} \stackrel{\kappa}{=} D_{T}\} \ll 0.5$. In summary, this means that the assumption of $\kappa$-cover is not fulfilled if the interest lies in particular in the center of the distributions, but is fulfilled if this requirement is sufficiently reduced.\\

When applying the C-test, $\kappa$ has always to be chosen beforehand, since it determines the minimum extent of covering one is willing to accept. One could, e.g., adopt the $6\sigma$-approach, i.e., the respective $\kappa$ is chosen to be $1-\Phi(3)=0.00135$. Depending on the criticality relevant to clinical outcome, larger values such as 0.01 or 0.05 should be selected. The higher the criticality of the QA in question is classified, meaning larger $\kappa$, the smaller is the scope for deviations of $D_T$ from $D_R$. In order to illustrate the comparison of $D_T$ and $D_R$ by applying the C-test, different choices of $\kappa$ and $\alpha_p$ are used.\\

The empirical quantiles of both distributions can be found in panel a) of Table \ref{indquant}. It can be seen that in all cases $\hat{Q}^{1-\kappa}_R < \hat{Q}^{1-\kappa}_T$ and the question is whether this deviation is statistically significant. Applying the $C$-test yields the results (p-values) shown in panel b) Table \ref{indquant}. These p-values indicate that the null-hypothesis of $D_R$ being a $\kappa$-cover of $D_T$ cannot be rejected for all ${\kappa}$ which are at least slightly smaller than 0.1. Therefore, it seems reasonable to assume that the above discussed borderline value of $\kappa$ lies around 0.1. The values of $K$ necessary for applying the $C$-test to samples of 24 and 48 observations respectively, have been determined by simulations. However, they are in fact less relevant in this case, since for $q \leq 0.5$ we have that $Q^{q}_R \ll Q^{q}_T$ and therefore the pretest significantly rejects $H_{0l}$ for all chosen values of $\alpha_p$. Hence, based on the pre-test we effectively use a one-sided test.\\  

\begin{table}[]

\centering

\caption{\label{indquant} Empirical quantiles and p-values of the C-test}

\vspace*{1\baselineskip}

\setlength\extrarowheight{-3pt}

\begin{tabular}{cc}

\bigskip
\centering
\begin{tabular}{rrr}
\multicolumn{3} {c} {a.) Empirical quantiles}\\   
\multicolumn{3} {c} {of $D_T$ and $D_R$}\\
  \hline\\[-4pt]
  $1-\kappa$ & $\hat{Q}^{1-\kappa}_R$ & $\hat{Q}^{1-\kappa}_T$\\[5pt] 
  \hline
0,8 &	213,6 &	233,4 \\ 
0,9 &	223 &	240,4 \\ 
0,95 &	223 &	245,54 \\ 
0,9987 &	251,1 &	269,7 \\ 
 \hline
\end{tabular}


\bigskip
\centering
\begin{tabular}{rrrrrr}
 & & & & &     \\
\end{tabular}


\bigskip
\centering
\begin{tabular}{rrrrr}    
\multicolumn{5} {c} {b.) p-values from applying the}\\
\multicolumn{5} {c} {C-test to the data set from industry}\\
\hline
\backslashbox{$\kappa$}{$\alpha_p$} 
&\makebox[3em]{0.01}&\makebox[3em]{0.05}&\makebox[3em]{0.1}
&\makebox[3em]{0.2}\\\hline
0.2 &	0.0103 &	0.0119 &	0.0091 &	0.0134\\
0.1 &	0.0499 &	0.0527 &	0.0501 &	0.0462\\
0.05 &	0.112 &	0.1095 &	0.1094 &	0.1115\\
0.00135 &	0.441 &	0.4258 &	0.4403 &	0.4334\\
\hline
\end{tabular}
\end{tabular}

\end{table}

To summarize, one can see that given a significance level of 0.05 for the $C$-test, the assumption of analytical similarity of the test product to the originator cannot be rejected with regards to the QA in question for all $\kappa < 0.1$ despite the fact that for the empirical quantiles we have that $\hat{Q}^{1-\kappa}_R < \hat{Q}^{1-\kappa}_T$ for all $\kappa$ under investigation. The result of the statistical analysis corresponds well to the subject-matter experts opinion which initially classified the product as borderline but ultimately assessed it as permissible.\\

This example illustrates well the importance of the choice of $\kappa$, which depends on the criticality of the QA under investigation. For a QA which has low to moderate relevance, a $\kappa<0.1$ should be fine, otherwise a $\kappa>0.1$ should be chosen, which would result in rejection of the product at hand.\\

\section{Summary \& Outlook}

The concept of $\kappa$-cover is introduced for comparing distributions with respect to the needs in several fields of industrial production. The intended application is for the special case of comparing QAs of a test product (e.g. generic drug) to a reference product (e.g. originator) when there exists no explicit knowledge about accepted ranges and applicable tolerances. Depending on the criticality of the QA the amount of deviation considered acceptable between the distributions of the QA of the test and the reference product varies. The concept presented in this work allows quantification of the minimum level to which the range of the distribution of the QA of the reference product has to cover the range of the distribution of the QA of the test product. For the comparison task at hand, suitable statistical hypotheses are formulated and a statistical test for normally distributed QAs is constructed. The proposed test extends a test based on the concept of generalized p-values for the comparison of two quantiles to the situation of simultaneously and adaptively testing in the upper and lower tails of the distributions. This novel test has favorable finite sample properties in terms of size and power as demonstrated by our simulations, although it is slightly undersized for small sample sizes at certain parameter constellations. This problem vanishes with increasing sample size, i.e., the test is asymptotically unbiased. However, the extremely small samples that are often available for the application of such tests may not be sufficient to ensure adequate testing power. Therefore, irrespective of the method applied, it is recommended to carry out simulations to investigate the $\beta$ error as a function of the parameters to be selected, e.g. $\kappa$, and the sample size in order to be able to correctly assess the result of the statistical comparison. At least from a consumer perspective, it would of course be preferable, to reverse the statistical hypotheses of the comparison task such that analytical similarity is stated in the alternative hypothesis. This is, as outlined above, only reasonable, if tolerance can be granted. Application of tolerance together with the reversal of the hypotheses in the context of $\kappa$-cover and in particular the construction of a suitable test which properly deals with the multiple testing problem that, as discussed in detail above, inevitably emerges when two or more simultaneous quantile comparisons are performed, are currently investigated. The results are promising, especially it is again possible to construct a test that shows the desired asymptotic behavior. Nevertheless, it remains questionable whether more powerful tests can be constructed for range based comparisons, in particular without relying on the assumption of normally distributed QAs. At the moment, especially if no tolerance can be granted, it seems that with small sample sizes, one has to sacrifice either consumer safety or producer interests. \\


\noindent \textbf{CONFLICT OF INTEREST}\\
\noindent The authors declare that they have no conflicts of interest for this work.\\

\noindent \textbf{DATA AVAILABILITY STATEMENT}\\
\noindent Data sharing is not applicable to this article as no new data were created or analyzed in this study.



\begin{thebibliography}{10}
\newcommand{\printfirst}[2]{#1}
\newcommand{\switchargs}[2]{#2#1}
\providecommand{\url}[1]{\normalfont{#1}}
\providecommand{\urlprefix}{Available at }

\bibitem{EMA.2021}
 EMA, \emph{Reflection paper on statistical methodology for the comparative
  assessment of quality attributes in drug development},
  \url{https://www.ema.europa.eu/en/statistical-methodology-comparative-assessment-quality-attributes-drug-development#current-version-section}
  (2021). Accessed: 10 August 2021.


\bibitem{Schuirmann.1987}
D.J. Schuirmann, \emph{A comparison of the two one-sided tests procedure and
  the power approach for assessing the equivalence of average bioavailability},
  Journal of Pharmacokinetics and Biopharmaceutics 15 (1987), p. 657-680.


\bibitem{Mielke.2018}
J. Mielke, F. Innerbichler, M. Schiestl, N.M. Ballarini, and B. Jones,
  \emph{The assessment of quality attributes for biosimilars: a statistical
  perspective on current practice and a proposal}, The AAPS journal 21 (2018),
  p.~7.



\bibitem{ChungChow.2014}
S. {Chung Chow}, \emph{On assessment of analytical similarity in biosimilar
  studies}, Drug Designing: Open Access 03 (2014).



\bibitem{FDA.2017}
 FDA, \emph{Statistical approaches to evaluate analytical similarity guidance
  for industry} (2017).

\bibitem{FDA.2018}
 FDA, \emph{Fda withdraws draft guidance for industry: Statistical approaches
  to evaluate analytical similarity},
  \url{https://www.fda.gov/drugs/drug-safety-and-availability/fda-withdraws-draft-guidance-industry-statistical-approaches-evaluate-analytical-similarity}
  (2018). Accessed: 15 July 2021.


\bibitem{Weng.2019}
Y.T. Weng, Y. Tsong, M. Shen, and C. Wang, \emph{A modified wald test for
  reference scaled assessment of analytical equivalence}, Journal of
  Biopharmaceutical Statistics 29 (2019), pp. 1068--1081.


\bibitem{FDA.2019}
 FDA, \emph{Development of therapeutic protein biosimilars: Comparative
  analytical assessment and other quality-related considerations guidance for
  industry},
  \url{https://www.fda.gov/regulatory-information/search-fda-guidance-documents/development-therapeutic-protein-biosimilars-comparative-analytical-assessment-and-other-quality}
  (2019). Accessed: 10 August 2021.


\bibitem{Tsong.2017}
Y. Tsong, X. Dong, and M. Shen, \emph{Development of statistical methods for
  analytical similarity assessment}, Journal of Biopharmaceutical Statistics 27
  (2017), pp. 197--205.


\bibitem{Chow.2016}
S.C. Chow, F. Song, and H. Bai, \emph{Analytical similarity assessment in
  biosimilar studies}, The AAPS journal 18 (2016), pp. 670--677.

\bibitem{Wang.2017}
T. Wang and S.C. Chow, \emph{On the establishment of equivalence acceptance
  criterion in analytical similarity assessment}, Journal of biopharmaceutical
  statistics 27 (2017), pp. 206--212.

\bibitem{Son.2020}
S. Son, M. Oh, M. Choo, S.C. Chow, and S.J. Lee, \emph{Some thoughts on the qr
  method for analytical similarity evaluation}, Journal of biopharmaceutical
  statistics 30 (2020), pp. 521--536.



\bibitem{Chow.2021}
S.C. Chow and S.J. Lee, \emph{Current issues in analytical similarity
  assessment}, Statistics in Biopharmaceutical Research 13 (2021), pp.
  203--209.


\bibitem{Boulanger.2016}
B. Boulanger, \emph{Assessment of analytical biosimilarity: the objective, the
  challenge and the opportunities}, \url{https://www.efspi.org/Documents/E v e
  n t s / R e g u l a t o r y % 2 0 M e e t i n g s / 2 0 1 6
  /5.2.Bruno%20Boulanger_EFSPI_talk_analytical_biosimilarity_13SEP2016_V3.pdf}
  (2016). Accessed: 10 August 2021.



\bibitem{Jarque.1987}
C. M. Jarque and A. K. Bera \emph{A test for normality of observations and regression residuals}, International Statistical Review. 55 (2): 163-172. doi:10.2307/1403192. JSTOR 1403192.

\bibitem{Shapiro.1965}
S. S. Shapiro and M. B. Wilk \emph{An analysis of variance test for normality (complete samples)}, Biometrika. 52 (3-4): 591-611. doi:10.1093/biomet/52.3-4.591. JSTOR 2333709. MR 0205384. p. 593

\bibitem{Neyman.1937}
J. Neyman \emph{"Smooth test" for goodness of goodness of fit}, Skandinaviske Aktuarietidskrift 20:150-199, 1937.





\bibitem{Holtzman.1950}
W.H. Holtzman, \emph{The unbiased estimate of the population variance and
  standard deviation}, The American Journal of Psychology 63 (1950), p. 615.

\bibitem{Miller.1981}
R.G. Miller, \emph{Simultaneous Statistical Inference}, second edition ed.,
  Springer eBook Collection Mathematics and Statistics, {Springer New York},
  New York, NY, 1981.

\bibitem{Guo.2005}
H. Guo and K. Krishnamoorthy, \emph{Comparison between two quantiles: The
  normal and exponential cases}, Communications in Statistics - Simulation and
  Computation 34 (2005), pp. 243--252.

\bibitem{Tsui.1989}
K.W. Tsui and S. Weerahandi, \emph{Generalized p-values in significance testing
  of hypotheses in the presence of nuisance parameters}, Journal of the
  American Statistical Association 84 (1989), p. 602.

\bibitem{Weerahandi.1995}
S. Weerahandi, \emph{Exact Statistical Methods for Data Analysis}, Springer
  eBook Collection Mathematics and Statistics, Springer, New York, NY, 1995.


\bibitem{Hannig.2016}
Hannig, Jan and Iyer, Hari and Lai, Randy C. S. and Lee, Thomas C. M., \emph{Generalized Fiducial Inference: A Review and New Results}, 
Journal of the American Statistical Association (2016), pp. 1346--1361.


\bibitem{Murph.2023}
Murph, Alexander C. and Hannig, Jan and Williams, Jonathan P., \emph{Introduction to Generalized Fiducial Inference}, 
arXiv (2023), pp. 1346--1361, doi = {10.48550/arXiv.2302.14598}



\bibitem{Stangler.2019}
Stangler, Thomas and Schiestl, Martin, \emph{Similarity assessment of quality attributes of biological medicines: the calculation of operating characteristics to compare different statistical approaches},
AAPS Open 5 (1) (2019).


\bibitem{Niazi.2022}
Niazi, Sarfaraz K., \emph{Biosimilars: Harmonizing the Approval Guidelines.}, Biologics 2022, 2, 171-195. https://doi.org/10.3390/biologics2030014 



\end{thebibliography}



\begin{appendices}


%

\section{Calculation of the generalized $p$-values for the tests based on $T_3$} \label{App 1}

Everything stated in Guo \& Krishnamoorthy\cite{Guo.2005} for $T_2$ also holds for $T_3$. Since, for a given $(\bar{x}_{R}, \bar{x}_{T}, s^{2}_{R}, s^{2}_{T})$, the distribution of $T^u_3$ is independent of any unknown parameters (see second equality of \ref{eq1}), this distribution is accessible by simulation and a test for the hypotheses in (Hyp2) can be based on $T^u_3$. The generalized $p$-value for the corresponding test is given by $p_u=P(T^u_3 > 0)$ (Guo \& Krishnamoorthy\cite{Guo.2005} use the reverse order of the quantities in (T3) in calculating $T_2$ in which case the p-value is given by $P(T_2<0)$.) and analogously to Guo \& Krishnamoorthy\cite{Guo.2005}, the test based on $T^u_3$ rejects the null hypothesis of (Hyp 2) whenever 

\vspace*{-1\baselineskip}

\begin{equation} \label{gpv} \tag{gpv} p_u=P(T^u_3 > 0) < \alpha, \end{equation}

\vspace*{.5\baselineskip}

\noindent with $\alpha$ denoting the chosen level of significance. The generalized $p$-value can be estimated using simulations. As proposed by Guo \& Krishnamoorty, $T^u_3$ is computed for $N$ (e.g. $N=10^5$) simulated vectors $(Z, U^{2}_{R}, U^{2}_T)$ and the generalized p-value in (gpv) is estimated by the proportion of the values of $T^u_3$ which are greater than zero.\\

Also in the case of $T^l_3$ and $T_{|3|}$, the corresponding p-value is again computed by the proportion of simulated draws being greater than zero, i.e., \label{pl} $p_l=P(T_3^l>0)$ and \label{p3} $p_{|3|}=P(T_{|3|}>0)$.\\

When estimating the respective generalized p-values of all tests involved in the calculation of  $p_C$ by simulation, as explained above, the same random numbers, i.e., $(Z, U^{2}_{R}$ and $U^{2}_{T})$, generated anew in each iteration, are used for computing the values of all test statistics combined in the $C$-test. This means that it is not necessary to generate different random vectors $(Z, U^{2}_{R}, U^{2}_{T})$ for the different tests in each iteration, but that a single newly generated random vector per iteration is sufficient.


\section{Asymptotic behavior of the $C$-test} \label{App 2}

To better understand how the test works and that it is asymptotically justified, consider the following two cases:

Case 1: Assume $D_T\neq D_R$ and that, w.l.o.g., $Q^R_{\kappa}=Q^T_{\kappa}$ and $Q^R_{1-\kappa}>Q^T_{1-\kappa}$. In this case the lower quantile lies on the boundary of $H_0$, but the upper quantile of $D_R$ is strictly larger than that of $D_T$. Therefore, the auxiliary hypothesis $H_{0l}$ holds, but $H_{0u}$ is violated. Due to the consistency of the test it holds that $\bar{p}_u \rightarrow 0$ as $min(n_1,n_2)\rightarrow \infty$. Conversely, $p_u \rightarrow 1$. Hence, asymptotically $\bar{p}_{min} \rightarrow 0$ and the unadjusted test is applied, i.e., the appropriate one-sided test (recall that $P(T_{|3|}>0)=min(P(T^u_3>0), P(T^l_3>0))$).

Case 2: Assume that $D_T= D_R$. In this case, irrespective of the sample size, one can distinguish several sub-cases: none of the pretests is significant, only one of them is significant or both are significant simultaneously. In all, but the last sub-case, also the respective test in the second step might be significant, i.e., can reject $H_0$ of (Hyp1). The only possibility to exert influence on the overall rejection probability of the test procedure is by choosing $K$ since the significance level of the one-sided tests in the second step is determined by the desired significance level  $\alpha$ of the C-test. As already mentioned, the tests involved are not independent of each other which makes an analytical treatment of this problem very challenging. By using simulations a value for $K$ can be determined such that the rejection probability of the C-test is equal to $\alpha$ if $D_T= D_R$.

\section{Monte Carlo Simulations} \label{App 3}

\begin{figure*}[h]
 \centering
 \includegraphics[width=16cm]{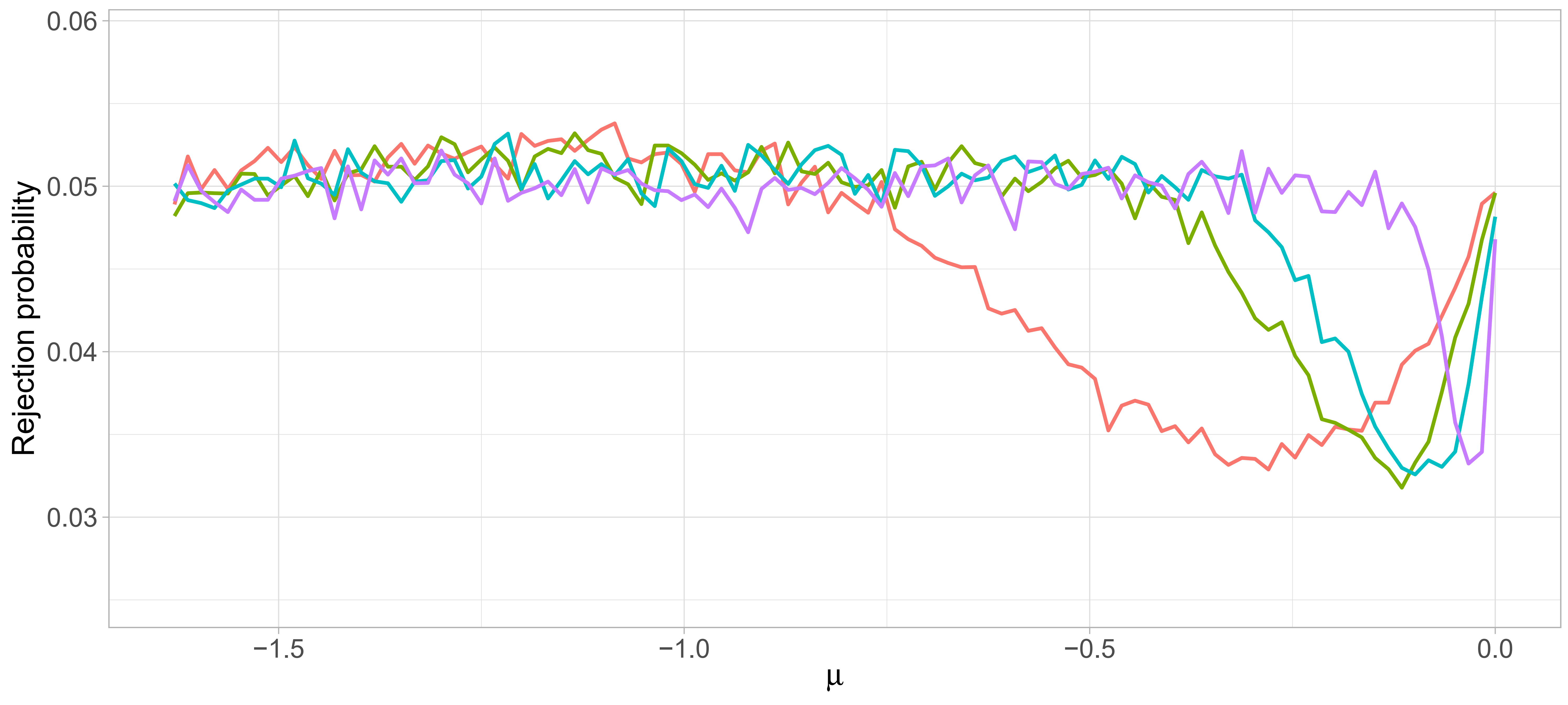}
 \caption{Rejection probability of the C-test for certain parameter combinations on the border of $\Theta_0^{\kappa}$. $D_R=N(0,1)$, $\kappa=0.05$ and $\alpha=0.05$. red: n=10, green: n=50, blue: n=100, violet: n=10000.} \label{Sim1}
\end{figure*}
\begin{figure}[h]

\centering

\subfloat[Subfigure 1 list of figures text][]{
\includegraphics[width=0.45\textwidth]{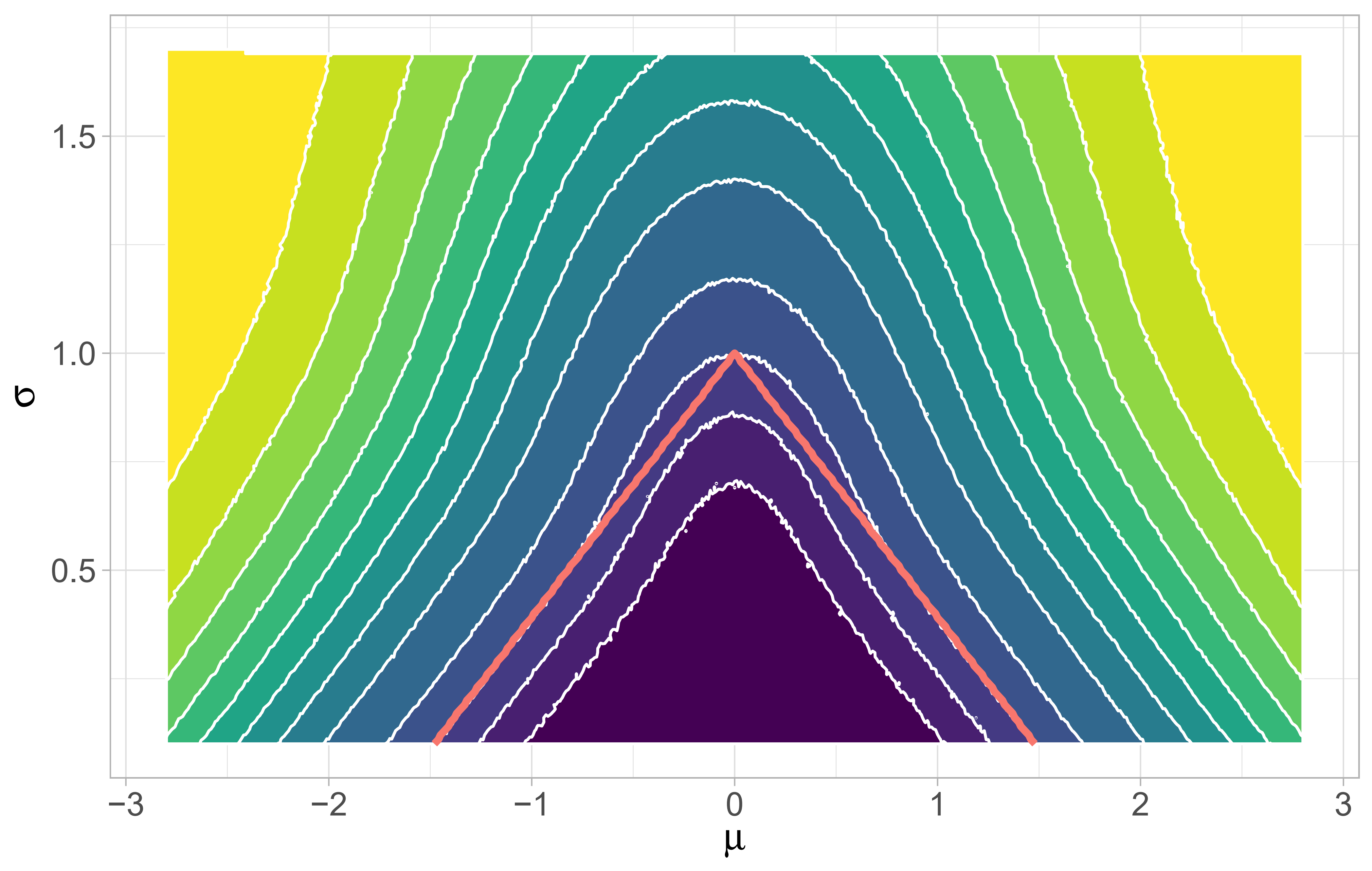} 
\label{fig:subfig1}}
\qquad
\subfloat[Subfigure 2 list of figures text][]{
\includegraphics[width=0.45\textwidth]{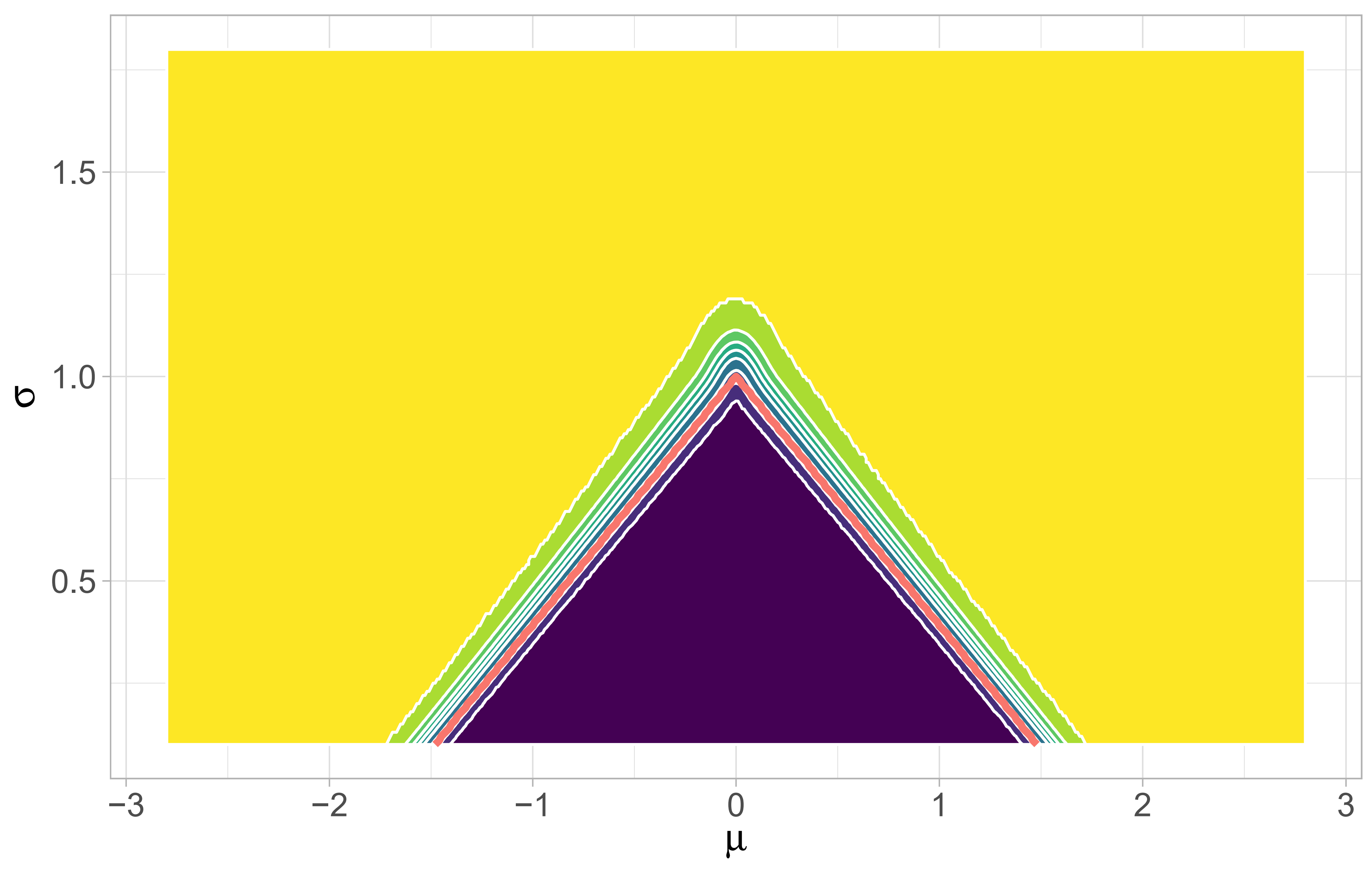}
\label{fig:subfig2}}

\caption{Rejection probability with respect to sample size. $D_R=N(0,1)$, $\kappa=0.05$ and $\alpha=0.05$. (a) n=10, (b) n=1000. The contour lines represent the following probabilities (increasing with increasing distance from the origin):  n=10: 1\%, 2.5\%, 5\%, 10\%, 20\%, 30\%, 40\%, 50\%, 60\%, 70\%, 80\% and 90\%, n=1000: 0.1\%, 2\%, 10\%, 30\%, 50\%, 70\%, 90\%, 100\% .The wavy contour lines in both plots are due to numerical artifacts.}\label{fig:contour}


\end{figure}

In addition to the simulation results given above in the tables \ref{Kor} - \ref{tab:power}, a lot more information regarding the simulation results is represented by Figures \ref{Sim1} and  \ref{fig:contour}.

Figure \ref{Sim1} shows some of the simulation results regarding the size of the $C$-test. The data shown in this figure was generated based on the following settings: $D_R=N(0,1)$ and the size of the C-test, $\alpha$, is chosen to be $0.05$. Since Figure \ref{Sim1} focuses on the border of $\Theta^{\kappa}_{(0,1),0}$, the respective x-interval is divided into a sequence with a step size of $0.01$ for choosing the varying parameter(s) of $D_T$. For each chosen parameter combination of $D_T$, $5 \cdot 10^4$ iterations are used to calculate the respective rejection probability with respect to the C-test procedure outlined above (each p-value is calculated based on $10^4$ random vectors). In all cases the size is maintained and the bias vanishes with increasing sample size, i.e., the proposed test is asymptotically unbiased.

The data shown in the subfigures of Figure \ref{fig:contour} was generated as follows: w.l.o.g., in both cases $D_R=N(0,1)$, $\kappa=0.05$, $\alpha_p=0.05$ and $\alpha$, the size of the C-test, is also chosen to be $0.05$. The parameters for $D_T$ are chosen with respect to a grid with a grid size of $0.01$ in every dimension. This grid was placed in $\Theta^{}_{(0,1)}$ over the respective $H_0$ parameter space $\Theta^{0.05}_{(0,1),0}$, such that it and also its vicinity is reasonably covered. For each chosen parameter combination of $D_T$, represented by the grid points, $5 \cdot 10^4$ iterations are used to calculate the respective rejection probability with respect to the C-test procedure outlined above (each p-value is calculated based on $10^4$ random vectors). The color scheme used in Figure \ref{fig:contour} depicts the rejection probability in the following way: it ranges from dark purple, representing a rejection probability of $0$, to a bright yellow, representing a rejection probability of $1$. The contour line going through the tip of the red triangle (equal distributions) indicates the desired level of significance (i.e., 0.05). The rejection probability is lowest inside the $H_0$ parameter space and is increasing with increasing distance from the point of origin. The red triangle indicates $\partial\Theta_0 \cap \Theta_0$ and it can be seen that the rejection probability is less than or equal to the desired level of significance for all $\theta \in \Theta_0$. Subfigure (a) depicts the rejection probability of the $C$-test for $n=10$ and subfigure (b) those for $n=1000$. For $n=1000$, the rejection probability is zero almost everywhere inside and 1 outside $\Theta_0^{\kappa}$, and increases rapidly inside the narrow band along the border of $\Theta_0^{\kappa}$. Visual inspection alone can, as opposed to (a), no longer detect bias. Figure \ref{Sim1} gives a more detailed picture regarding the behavior of the C-test on $\partial\Theta_0$ and it shows in particular that the bias vanishes with increasing sample size.



\end{appendices}

\end{document}